\documentclass[a4paper,twocolumn,11pt,accepted=2024-07-07]{quantumarticle}
\pdfoutput=1

\usepackage{bbm}
\usepackage{mathrsfs}
\usepackage{epsfig}

\usepackage[numbers,sort&compress]{natbib}

\usepackage{soul,color}
\usepackage{graphicx}

\usepackage{amsfonts}
\usepackage{amsthm}
\usepackage[figuresright]{rotating}
\usepackage{amssymb}
\usepackage{amsmath}
\usepackage{dcolumn}
\usepackage{physics}
\usepackage{float}
\usepackage{bm}
\usepackage{verbatim}
\usepackage{braket}
\usepackage{babel}
\usepackage{multirow}
\usepackage[normalem]{ulem}
\usepackage[colorlinks,linkcolor=blue,anchorcolor=blue,citecolor=blue,urlcolor=blue]{hyperref}
\usepackage{stackengine}
\usepackage{qcircuit}
\usepackage{tabularx}

\newcounter{protocol}
\newenvironment{protocol}[1]
  {\par\addvspace{\topsep}
   \noindent
   \tabularx{\linewidth}{@{} X @{}}
    \hline
    \refstepcounter{protocol}\textbf{Protocol \theprotocol} #1 \\
    \hline}
  { \\
    \hline
   \endtabularx
   \par\addvspace{\topsep}}

\newtheorem{theorem}{Theorem}
\newtheorem{lemma}[theorem]{Lemma}
\newtheorem{definition}{Definition}
\newtheorem*{problem*}{Main Problem}
\usepackage{fancyhdr}

\newcommand{\sinc}{\text{sinc}}
\newcommand{\sign}{\text{sign}}
\newcommand{\nn}{\nonumber}

\usepackage{orcidlink}

\begin{document}
  
\title{Single-shot Quantum Signal Processing Interferometry}

\author{Jasmine Sinanan-Singh}\thanks{These two authors contributed equally.}
 \affiliation{Department of Physics, Co-Design Center for Quantum Advantage, Massachusetts Institute of Technology, Cambridge, Massachusetts 02139, USA}

\author{Gabriel L. Mintzer \orcidlink{0000-0002-5985-9958}}\thanks{These two authors contributed equally.}
 \affiliation{Department of Physics, Massachusetts Institute of Technology, Cambridge, Massachusetts 02139, USA}
 \affiliation{Department of Electrical Engineering and Computer Science, Massachusetts Institute of Technology, Cambridge, Massachusetts 02139, USA}

\author{Isaac L. Chuang}
 \affiliation{Department of Physics, Co-Design Center for Quantum Advantage, Massachusetts Institute of Technology, Cambridge, Massachusetts 02139, USA}
 \affiliation{Department of Electrical Engineering and Computer Science, Massachusetts Institute of Technology, Cambridge, Massachusetts 02139, USA}

\author{Yuan Liu\,\orcidlink{0000-0003-1468-942X}}\email{q\_yuanliu@ncsu.edu}
 \affiliation{Department of Physics, Co-Design Center for Quantum Advantage, Massachusetts Institute of Technology, Cambridge, Massachusetts 02139, USA} 
 \affiliation{Department of Electrical and Computer Engineering, North Carolina State University, Raleigh, North Carolina, 27606, USA}
 \affiliation{Department of Computer Science, North Carolina State University, Raleigh, North Carolina, 27606, USA}
 \affiliation{Department of Physics, North Carolina State University, Raleigh, North Carolina, 27606, USA}

\begin{abstract}
    Quantum systems of infinite dimension, such as bosonic oscillators, provide vast resources for quantum sensing. Yet, a general theory on how to manipulate such bosonic modes for sensing beyond parameter estimation is unknown. We present a general algorithmic framework, quantum signal processing interferometry (QSPI), for quantum sensing at the fundamental limits of quantum mechanics by generalizing Ramsey-type interferometry. Our QSPI sensing protocol relies on performing nonlinear polynomial transformations on the oscillator's quadrature operators by generalizing quantum signal processing (QSP) from qubits to hybrid qubit-oscillator systems. We use our QSPI sensing framework to make efficient binary decisions on a displacement channel in the single-shot limit. Theoretical analysis suggests the sensing accuracy, given a single-shot qubit measurement, scales inversely with the sensing time or circuit depth of the algorithm. We further concatenate a series of such binary decisions to perform parameter estimation in a bit-by-bit fashion. Numerical simulations are performed to support these statements. Our QSPI protocol offers a unified framework for quantum sensing using continuous-variable bosonic systems beyond parameter estimation and establishes a promising avenue toward efficient and scalable quantum control and quantum sensing schemes beyond the NISQ era.
\end{abstract}

\maketitle

\tableofcontents

\section{Introduction}
\label{sec:intro}

Sensing and metrology are fundamental pursuits of science and technology, and quantum systems have been used to advance metrological precision to new bounds \cite{ligo2016observation,bothwell2022resolving,roussy2023improved,giovannetti2004science}. Typical quantum sensing protocols involve manipulation of quantum coherence and entanglement followed by measurement to extract useful classical information from quantum systems \cite{degen2017quantum,ye2024essay}. 

The efficiency of different quantum sensing protocols varies by construction. At a high level, any sensing protocol can be assessed by the space and time resource requirements (e.g., the size of the quantum sensor, the length of the sensing protocol, any required repetition of the experiments) that it needs in order to achieve a given sensitivity in the sensing task, for example estimating a given parameter to a certain precision. The intrinsically probabilistic nature of quantum systems necessarily introduces uncertainty into the measurement result of any sensing protocol, leading to the so-called standard quantum limit (SQL). In the SQL, the standard deviation of the estimated parameter scales inversely as the square-root of the space and time resources employed, as is familiar in processes involving shot noise.

By leveraging non-classical properties of quantum states like entanglement \cite{boixo2008quantum,tilma_entanglement_2010} and general quantum correlations \cite{braun_quantum-enhanced_2018} or using coherent sampling of the signal and adaptive feedback \cite{Berry_2000,G_recki_2020,marciniak_optimal_2022}, sensitivity in parameter estimation can be improved beyond the SQL to approach a more fundamental physical limit, the Heisenberg limit (HL). The HL dictates that the scaling of precision with total sensing time $t$ can be no better than $1/t$; equivalently, with $N$ probes used in an experiment, the precision scales no better than $1 / N$. 

In fact, this fundamental physical limit has been achieved by a number of sensing protocols.  One of the oldest and best-known is the interferometric protocol known as cat-state sensing, named after Schr{\"o}dinger's cat for its use of superpositions of two distinct macroscopic states, such as the all-spin-up and all-spin-down states in a multi-atom system. This protocol, first realized for spin-states in 1996 by Bollinger \textit{et al.}, achieves the optimal bound for frequency uncertainty of an $N$-particle system \cite{catstateinterferometry}.  
This optimal HL bound, equal to $(NT)^{-1}$, where $T$ is the time for a single repetition of the protocol, is achieved by modifying the Ramsey technique \cite{RamseyNormanF.1950AMBR} to use a maximally correlated GHZ state and a different final measurement operator. 
As it achieves the HL, this variety of cat-state sensing for spin systems has found broad application in precision phase sensing for atomic clocks \cite{marciniak_optimal_2022,Kaubruegger_2021}, where variational quantum algorithms are incorporated into multi-qubit Ramsey interferometry to iteratively optimize the sensing precision.

However, other parameters of interest, such as electric fields \cite{descamps_quantum_2023}, can be better sensed by bosonic modes (e.g., photonic and phononic oscillators) than by spin systems.
Bosonic sensors have been employed to perform precision sensing of small displacements to bosonic oscillators, and there have been many advances investigating the advantages of utilizing bosonic resource states \cite{penasa_disp_sensing,duivenvoorden_single-mode_2017}. Gilmore \textit{et al.} have found that coupling the spins of a trapped-ion crystal to their collective motional mode offers sub-SQL sensing performance, with the quantum enhancement achieved through interferometry of highly-entangled spin-motion cat states \cite{gilmore2021quantum}. Using the interference between squeezed light, the Advanced LIGO \cite{ligo2015squeezing} experiment can detect the space-time curvature changes induced by gravitational waves. Interferometric phase estimation using entanglement, coherent sampling, and adaptive feedback approaches the exact Heisenberg limit \cite{Daryanoosh_2018}, and similar results have been experimentally found using just a single bosonic mode \cite{Wang_2019}.

Additionally, coupling bosonic modes to other degrees of freedom can transfer information from one subsystem to another in order to facilitate more convenient measurement than the direct measurement of the bosonic modes themselves \cite{gietka_harnessing_2022}.
In spectroscopy, entangled cat-state laser sources have been used to enhance signals by an order of magnitude \cite{catstatespectroscopy}. 
Beyond the single-mode case, it has been demonstrated that the entanglement of many modes can provide HL-sensing enhancement for parameter estimation \cite{zhuang_distributed_2018,kwon2022quantum}. Furthermore, various efficient HL-scaling Hamiltonian learning protocols have been proposed, including some on bosonic systems, which can be rephrased as multi-parameter estimation problems \cite{li2023heisenberglimited,huang2023learning}.

Beyond parameter estimation, there are many other sensing applications that have been left largely unexplored---for example, single-shot decision-making. For such decision-making problems, the underlying signal can happen rarely, such as in the case of gravitational-wave detection \cite{ligo2016observation}, and it is therefore crucial to obtain useful information in the single-shot limit. When events are rare, many iterative protocols for parameter estimation and learning \cite{dong_beyond_2022,huang2023learning,zhou2018achieving,sugiura2023power,rossi2021quantum,rossi2022quantum,G_recki_2020} are challenged. Protocols for discrete decision problems such as classification on multiple bosonic modes have been developed in Ref.~\cite{zhuang2019physical,zhuang2021quantum} by using variational algorithms for state preparation and signal decoding with notable performance gain enabled by multi-mode entanglement. Despite the success of bosonic systems and cat-state sensing for parameter estimation, a unified protocol for general sensing tasks with provable speedup is unknown, particularly in scenarios where decisions must be made in the single-shot limit. Though there are many broad results concerning the optimal bounds on precision in quantum channel discrimination problems, there are few analyses of resource scaling in the single sample regime \cite{meyer2023quantum, q_illumination, qcd_pirandolo}. 

Protocols for realizing such general sensing tasks should generally build on the ability to perform transformations of the underlying signal. Not surprisingly, transformation of classical signals has been extensively studied in the context of \emph{signal processing} in electrical engineering \cite{oppenheim1999discrete}, where state-of-the-art classical algorithms have been developed to design a variety of \emph{filters} that transform the underlying classical signals in manners tailored to the desired purposes \cite{parks1972chebyshev,antoniou2018digital}. Inspired by classical signal processing, quantum signal processing (QSP) algorithms \cite{low2016methodology,low2019hamiltonian,kikuchi2023realization,rossi2022multivariable,dong2022infinite,rossi2023quantum,motlagh2023generalized,laneve2023quantum,rossi2023modular,martyn2023efficient,wang2022energy,MRTC21,yu2022power} can achieve arbitrary polynomial transformations on one or more quantum amplitudes. 
Given the triumphs of classical and quantum signal processing, might it be possible to adopt the philosophy of filter design to bosonic systems such that quantum signals on oscillators can be transformed for general sensing tasks?

\subsection{Contributions}
In the present work, we develop a novel algorithmic protocol for general quantum sensing tasks beyond parameter estimation using interferometric bosonic modes in a manner that enables \emph{systematic and analytically predictive improvement} of single-shot decision error, beyond what is possible with traditional sensing protocols. 
This \emph{QSP interferometry (QSPI)} protocol builds on a theory of bosonic QSP that can perform polynomial transformation on the block-encoded quadrature operators of bosonic modes using qubit rotations and qubit-oscillator entangling gates \cite{haljan2005spin,eickbusch2022fast}. The core of our QSPI protocol lies in these polynomial transformations generating nonclassical resource states for interferometery. Just as in typical Ramsey experiments \cite{RamseyNormanF.1950AMBR}, the signal being sensed in a QSPI experiment is queried only once, and the power of quantum enhancement comes from generating a non-classical resource state by increasing the QSP circuit length. The feature of querying the signal only once distinguishes our work from much prior art~\cite{dong_beyond_2022,huang2023learning,li2023heisenberglimited,Berry_2000}. The single-sample feature of our protocol (i.e., the property that only one measurement is needed) further distinguishes our work from \cite{meyer2023quantum,qcd_pirandolo}.

We illustrate the performance of the QSPI protocol with a theoretical analysis demonstrating its optimal extraction of binary decision information about a quantum displacement channel, which allows for improved scaling compared to classical protocols (see Def.~\ref{def:hl-decision}). As a concrete pedagogical example, we focus on the task of distinguishing or deciding whether a displacement channel has a displacement magnitude above or below a given threshold. 
This framework for quantum channel discrimination (QCD) problems opens the avenue to asking more complicated QCD questions that can decide between multiple hypotheses at the same time. 

Note that such decision problems are ideal for qubit-oscillator systems because we desire a single-shot measurement that answers a question about the channel acting on a bosonic quantum state with high probability. Given that the qubit is naturally binary under classical projective measurement, extracting a yes/no answer from the qubit should be much faster than extracting a continuous-valued answer by measuring the oscillator. 
This intuition is satisfied by our construction.
We also show how to utilize protocols for such simple decision problems to perform more complex tasks such as bit-by-bit estimation of the magnitude of a displacement.

The presentation below is organized as follows. We begin in Sec.~\ref{sec:binary-bosonic-decision} with a formal statement of the problem scenario and the expected performance using traditional cat-state sensing, then build on this to exhibit the new quantum signal processing interferometry protocol and its provably improved performance in Sec.~\ref{sec:alg-q-sense}. 
Building upon this, Sec.~\ref{sec:thres-binary-decision} analyzes analytically the performance of our QSPI protocol for making binary decisions about the displacement parameter.
Sec.~\ref{sec:quantum_decisions} then demonstrates the use of such a decision-making subroutine to perform parameter estimation on the magnitude of a displacement by combining with classical decision-making theory. Sec.~\ref{sec:numerical} presents numerical results, revealing an efficient scaling in the decision error and agreeing with analytical expectations. Conclusions and outlook are given in Sec.~\ref{sec:conclusion}.

\section{A Binary Bosonic Decision Problem}
\label{sec:binary-bosonic-decision}

In this section, we first set up some notation and define quantum decision-making problems in displacement sensing, as well as what \emph{efficient} behavior is for decision problems in Sec.~\ref{sec:quantum-decision-def}. A brief discussion of classical protocols for decisions is presented in Sec.~\ref{sec:classical-decision-protocols} to contrast with our quantum protocols. In Sec.~\ref{sec:cat-state-sensing}, we review the basics of a typical displacement-sensing protocol based on cat-state interferometry, highlighting its advantages and limitations in order to motivate why a more general sensing scheme is required.

\subsection{Quantum Decision-Making for a Displacement Channel}
\label{sec:quantum-decision-def}

We consider a quantum sensing problem in a joint qubit-oscillator system subjected to a unitary displacement channel $S_\beta$
\begin{align}
    S_\beta := \begin{bmatrix}
        e^{i \beta \hat{p}}  & 0 \\
        0 & e^{i \beta \hat{p}}
    \end{bmatrix},
    \label{s}
\end{align}
where we have written $S_\beta$ under the joint qubit-oscillator tensor product form such that $S_\beta = I \otimes e^{i \beta \hat{p}}$ and $\beta$ is the amount of the position kick acting on the oscillator;  $e^{i \beta \hat{p}} \ket{x}_{\rm osc} = \ket{x - \beta}_{\rm osc}$ for a position eigenstate $\ket{x}_{\rm osc}$ (the subscript ``osc'' refers to ``oscillator'' to distinguish it from the qubit register). The symbol $:=$ used here represents the definition of a quantity, and $\hat{x}, \,\hat{p}$ are the oscillator's canonical position and momentum operators, respectively.

On the joint system, we assume the resource gates are arbitrary single-qubit rotations $R_X(2 \theta) := e^{i\theta \hat{\sigma}_x}$ and a fixed qubit-oscillator entangling gate 
\begin{align}
    \mathcal{D}_c(i \kappa / \sqrt{2}) = e^{i \kappa \hat{x} \hat{\sigma}_z}
    \label{Dc-def}
\end{align}
parameterized by $\kappa$, where $\mathcal{D}_c(i \kappa / \sqrt{2})$ is a conditional displacement gate that imparts a momentum kick $\pm \kappa$ to the oscillator depending on the qubit state being $\ket{0}$ or $\ket{1}$. $\hat{\sigma}_x$, $\hat{\sigma}_z$ are the single-qubit Pauli matrices. This entangling gate is derived from the usual definition of a more general conditional displacement gate in phase space $\mathcal{D}_c(\alpha) := e^{\left(\alpha a^\dagger - \alpha^* a\right) \hat{\sigma}_z}$ by setting $\alpha = i \kappa / \sqrt{2}$. Moreover, the gate in Eq.~\eqref{Dc-def} is an operator with support on the infinite-dimensional qubit-oscillator joint Hilbert space. When acting on a position eigenstate of the oscillator $\ket{x}_{\rm osc}$, the gate given in Eq.~\eqref{Dc-def} reduces to $e^{i \kappa x \hat{\sigma}_z}$ which is simply a $2 \times 2$ operator acting on the qubit. Throughout the paper, we take $\hbar = 1$, $m = 1$, $\omega = 1$ for $m$ the mass of the oscillator and $\omega$ its angular frequency, in order to simplify our expressions. This means the fundamental length of the oscillator $\sqrt{\hbar / m \omega} = 1$; as a result, $\kappa$, which should take the unit of inverse $\beta$, also has a unit of 1. Additionally, we shall use $\hat{x}$, $\hat{p}$ and $x$, $p$ to distinguish the two different ways of using position and momentum as operators or real numbers. The product of $\kappa \hat{x}$ on the right-hand side of Eq.~\eqref{Dc-def} means the gate itself will be periodic in the oscillator position $x$ with a period of $T_x = \frac{2 \pi}{\kappa}$.

With these notations established, we are ready to define the quantum decision-making problem on the displacement channel:
\begin{problem*}[Quantum Binary Decision-Making for a Displacement Channel]
    Given $\beta_{\rm th} > 0$, construct a quantum circuit by using the resource gates $R_X$ and $\mathcal{D}_c(i \kappa / \sqrt{2})$ a maximum of $d$ times for some $\kappa$ to determine whether $|\beta| > \beta_{\rm th}$ or $|\beta| < \beta_{\rm th}$ with only a single query to $S_{\beta}$, such that the probability of making an erroneous decision, $p_{\rm err}$, is small.
\end{problem*}

Clearly, the probability $p_{\rm err}$ of erroneous decision  will depend on $\kappa$, $\beta_{\rm th}$, and $d$. Due to the periodicity of Eq.~\eqref{Dc-def} in $x$, any unitary constructed from repeated applications of $\mathcal{D}_c(i \kappa / \sqrt{2})$ and $R_X$ will be periodic in $x$ with the same period $T_x$. As will be discussed in Sec.~\ref{sec:qspi-displacement} (also see Fig.~\ref{fig:duality-cartoon}), the periodicity in $x$ for the $\mathcal{D}_c(\cdot)$ gate results in a period of $T_x / 2$ for $p_{\rm err}$ in terms of the sensing parameter $\beta$. This allows us to define a restricted region $\left(-\frac{\pi}{2\kappa}, \,\frac{\pi}{2\kappa}\right)$ where the sensing problem will be discussed. This notion of periodicity is similar to the concept of a ``unit cell'' in solid state physics \cite{ashcroft2022solid}.
Therefore, it is necessary to choose $\kappa$ to be small enough such that $\beta \in \left(-\frac{\pi}{2 \kappa}, \,\frac{\pi}{2 \kappa}\right)$ (in particular, $\beta_{\rm th}$ as well). However, $\kappa$ cannot be too small, or else $\mathcal{D}_c(i \kappa / \sqrt{2})$ will become too close to the identity operator, and its action on the qubit-oscillator system will not be effective. In the rest of the paper, we assume that $\kappa$ has been fixed with these conditions satisfied.
Furthermore, (as detailed in Appendix~\ref{app:perr-real-proof}) $p_{\rm err}$ is an even function of $\beta$; therefore, we only consider the case of $\beta_{\rm th} > 0$, as given in Main Problem.

Once we are given $\beta_{\rm th}$ and have fixed $\kappa$ as described above, it is instructive to consider how $p_{\rm err}$ scales as the number of resource gates $d$ in the sensing protocol. 
In the single-shot limit, since we are only allowed to query the signal $S_\beta$ once, it is not difficult to see that the most general single-shot decision-making protocol is as given in Fig.~\ref{fig:bosonic-QSP-sensing}a, where a state preparation routine is first used to prepare the joint qubit-oscillator system at some entangled quantum state, after which the signal of interest occurs to the oscillator. In the end, a signal decoding operation is applied to create some interference followed by a single-qubit measurement to extract the answer to the decision problem. Inspired by the definition of HL scaling in parameter estimation tasks, as discussed in Sec.~\ref{sec:intro}, we define \emph{efficient} scaling for the Main Problem:

\begin{definition}[Efficient Scaling for Binary Decision Error in the Main Problem.] 
    A sensing protocol achieves efficient scaling for binary decision-making with a displacement channel in the Main Problem if the resulting $p_{\rm err} \sim O(1 / d)$ up to a factor of $\text{polylog}(1 / d)$.
    \label{def:hl-decision}
\end{definition}

We use \emph{efficient} to term decision error scaling inversely with time, and we outline in the following section classical methods which are, in contrast, \emph{inefficient}, as their decision error scales inversely with the square root of time. We expect that there is a connection between the binary decision error scaling efficiently and achieving HL parameter estimation. We leave formal analysis of this connection for future work, but as our interferometric protocol provides a generalized way to move between local and global estimation, we suspect that HL scaling will arise from adaptive efficient extraction of binary information. Literature for quantum state and channel discrimination has placed broad bounds on the optimal error probability \cite{helstrom1969quantum,qcd_pirandolo,meyer2023quantum}, but these works are often lacking analysis of the resource requirements to achieve a given error.

\subsection{Classical Protocols for Decisions}
\label{sec:classical-decision-protocols}

Akin to defining the Heisenberg-limit for parameter estimation by comparing classical procedures, we can compare the scaling of our decision-making protocol to other classical and quantum methods. For single-shot decision-making, the classical analogue is akin to binary amplitude-shift keying (BASK) \cite{modulation} demodulation where a receiver must decide whether the vacuum coherent state $\ket{0}$ or some nominal coherent state $\ket{\alpha}$ with intensity $|\alpha|^2 = n$ has been sent. The source of error for classical receivers comes from the inherent overlap of coherent states, and thus classical procedures are fundamentally limited in distinguishing between coherent states \cite{helstrom1969quantum}. For BASK receivers, the signal-to-noise ratio increases as $\sqrt{t}$ for $t$ the signal integration time and thus limits the bit error rate to scale with $1 / \sqrt{t}$ \cite{modulation}.

By manipulating bosonic quantum states and an interferometric method, we can improve the scaling of error with time to $1/t$. In the following section, we will first consider simple supposition of coherent states as input sensing state prepared in the cat state sensing protocol (Fig.~\ref{fig:bosonic-QSP-sensing}b) to gain some intuition about interferometric scheme for single-shot decision-making protocol.

\subsection{Intuition from Cat-State Sensing}
\label{sec:cat-state-sensing}
\begin{figure}[htbp]
            \includegraphics[width=0.48\textwidth]{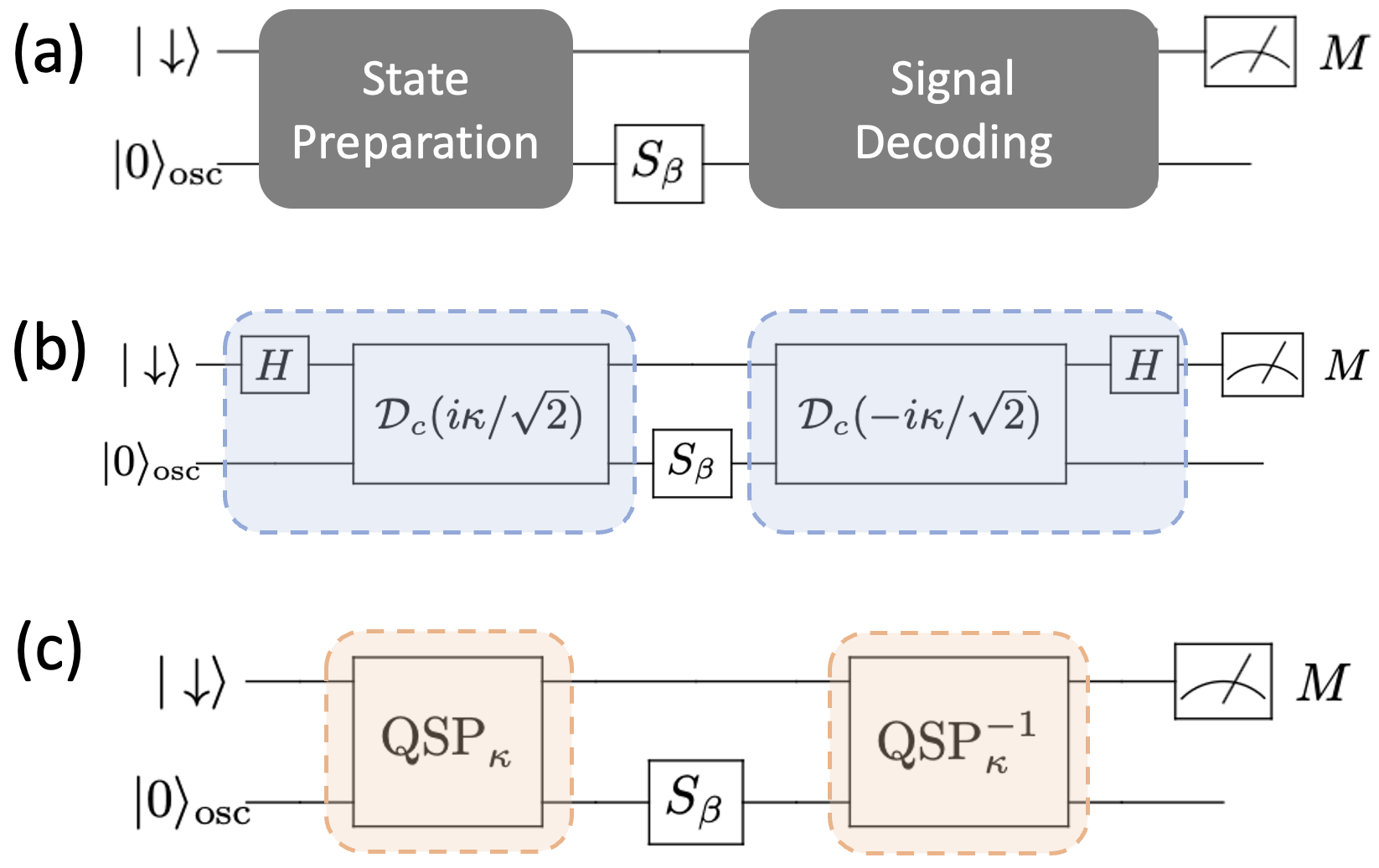}
            \caption{The most general single-shot decision-making protocol (a), and two realizations comparing the traditional cat-state sensing protocol (b) with the novel bosonic QSP interferometric protocol (c). In (c), the QSP operator creates an optimal sensing state, which then probes the signal $S_{\beta}$ and is finally un-created to produce desired interference, which is followed by a measurement on the qubit.
            }
            \label{fig:bosonic-QSP-sensing}
\end{figure}

The intuition for building a QSP interferometer comes from the cat-state protocol for sensing small displacements. A typical sensing scheme is shown in Fig.~\ref{fig:bosonic-QSP-sensing}b where a Hadamard gate and a controlled displacement $\mathcal{D}_c (i \kappa / \sqrt{2}) = e^{i \kappa \hat{x} \hat{\sigma}_z}$ are first used to prepare an entangled state of the qubit-oscillator joint system from an initial state of qubit at $\ket{\downarrow}$ and oscillator at vacuum $\ket{0}_{\rm osc}$ (first dashed blue box). The subscript $c$ in $\mathcal{D}_c(\cdot)$ means the displacement is controlled by the qubit. Then the underlying signal (a displacement ${S}_\beta = e^{i \beta \hat{p}}$) is applied to the oscillator and followed by another controlled displacement and a Hadamard gate (inverse of the previous dashed blue box). Finally, a qubit $Z$-basis measurement is performed. 

Using the commutation relationship between $\hat{x}$ and $\hat{p}$ we have the following relation
\begin{align}
    e^{i \kappa \hat{x}} e^{i \beta \hat{p}} = e^{i \beta \hat{p}} e^{i \kappa \hat{x}} e^{-i \kappa \beta},
    \label{commutation-x-p-exp}
\end{align}
which simplifies the final state of the joint qubit-oscillator system before the measurement to
\begin{align}
    \ket{\Psi_{\rm out}} = (\cos(\kappa \beta) \ket{\downarrow} + i \sin(\kappa \beta) \ket{\uparrow} ) \otimes e^{i \beta \hat{p}} \ket{0}_{\rm osc},
\end{align}
provided that the initial state is $\ket{\downarrow} \otimes \ket{0}_{\rm osc}$.
Therefore, the displacement $\beta$ is encoded in the amplitude of the ancilla qubit, where the measurement probability $p$ of the qubit in $\ket{\downarrow}$ is
\begin{align}
    p = {\rm Prob}[M = \ \downarrow] = \cos^2(\kappa \beta).
    \label{prob-cat-state-sensing}
\end{align}
Since $\kappa$ is known, we can repeat this sensing protocol multiple times in order to obtain an estimate of this probability, which will then tell us the value of $\beta$. 

More concretely, by repeating the protocol $N$ times, the standard deviation for estimating $\beta$ is given by $\Delta \beta$:
\begin{align}
    \Delta \beta = \frac{\Delta p}{\Big\lvert \frac{dp}{d \beta} \Big\rvert} = \frac{1}{2 \kappa \sqrt{N}}
    \label{cat-state-delta-beta}
\end{align}
where $\Delta p = \sqrt{\frac{p(1 - p)}{N}}$ is the standard deviation on $p$, estimated by performing $N$ experiments with a Bernoulli distribution, and the total time $t$ for repeating the sensing protocol $N$ times will be $t \propto N$.
An interesting observation immediately follows from Eq.~\eqref{cat-state-delta-beta}: for fixed $N$, $\Delta \beta$ improves roughly as $1 / \kappa$, where $\kappa$ is the displacement amount of the $\mathcal{D}_c (\cdot)$ gate. 
The physical intuition is that the cat state's small interference features in phase space have a characteristic length of $1 / \kappa$. As a result, the sensitivity on estimating a spatial variation in $\beta$ improves as $1 / \kappa$. It follows that taking $\kappa$ large would be beneficial to making high-sensitivity measurements.  Such large $\kappa$ can be realized in several ways, depending on the physical platform. For example, in trapped ions, a large $\kappa$ may be realized by increasing the laser pulse intensity or by increasing the pulse duration \cite{haljan2005spin}.

Despite its favorable sensing scaling in $\kappa$ for displacement sensing, the cat-state sensing protocol has some limitations. First, for fixed $\kappa$, $\Delta \beta$ decreases as $\frac{1}{\sqrt{N}}$ (or $\frac{1}{\sqrt{t}}$) as the number of classical repetitions $N$ increases. This is the typical shot noise statistical convergence rate corresponding to the  SQL and is sub-optimal as compared with HL scaling. Second, aside from parameter estimation on $\beta$, the cat-state sensing protocol is not particularly useful if we are only interested in learning partial information about the properties of $\beta$,  for example determining if $\beta$ is above or below a given threshold value $\beta_{\rm th}$. Since only partial information is needed in such scenarios, it is expected that more efficient sensing protocols exist. 

As alluded to earlier in the Introduction, filter designs in classical signal processing and advancement in quantum algorithms provide possibilities for overcoming the two aforementioned limitations such that: 1) more efficient (Heisenberg-limit) scaling can be achieved for parameter estimation, where $\Delta \beta \propto 1 / N$; and 2) the resulting sensing protocol works for other sensing tasks that only extract partial information about $\beta$.
The intuition is as follows: recall that the key feature of bosonic cat-state sensing is producing a fine-grained interference pattern in phase space that is sensitive to the displacement signal. If the $N$ incoherent repetitions of the cat-state sensing protocol can be concatenated together into a \emph{single-shot coherent protocol} which coherently manipulates the phase space interference pattern beyond that produced by the simple cat-state interferometer, then the coherent sensing state can be made sensitive to the partial information that we seek from the signal. In the next section, we give a construction of a novel QSP interferometer that circumvents the above two limitations and achieves efficient behavior for decision problems.

\section{Quantum Signal Processing Interferometry}
\label{sec:alg-q-sense}

Before presenting the QSPI construction, we first formulate a theory of bosonic QSP by alternating single-qubit rotations with controlled-displacement operations in Sec.~\ref{sec:bosonic-qsp} and show how to utilize this bosonic QSP approach as the basic building block to produce general sensing algorithms. Building upon the bosonic QSP theory for polynomial transformations of quadrature operators as well as the cat state sensing protocol above, we define and construct a novel QSP interferometer on hybrid qubit-oscillator platforms and present a new QSPI Theorem in Sec.~\ref{sec:qsp-interferometry}. Detailed analysis of the behavior of the QSP interferometer for a displacement operator is discussed in Sec.~\ref{sec:qspi-displacement}.

\subsection{Bosonic QSP Formalism}
\label{sec:bosonic-qsp}

Coupling a bosonic oscillator to a qubit is a useful approach for achieving universal control of the oscillator \cite{krastanov2015universal}. It has been shown that simple Jaynes-Cummings type interactions can achieve universal control on an arbitrary low-energy $d$-dimensional subspaces of an oscillator \cite{mischuck2013qudit,liu2021constructing}. By using an alternative dispersive coupling, universal control on oscillators has also been demonstrated using the echo-controlled displacement operator \cite{eickbusch2022fast}. Here, we draw a connection between these control protocols with quantum signal processing to develop a bosonic QSP formalism as a basic building block for the rest of the paper.

Quantum signal processing relies on two components: 1) a block-encoding of the signal operator; and 2) the ability to impart an arbitrary phase shift to the block-encoded operator. Block-encoding simply means embedding the target operator inside a known and accessible subspace of a unitary matrix. Methods for block-encoding on qubit devices are mostly limited to linear combination of unitaries \cite{childs2012hamiltonian,2015Berry}, and block-encoding of a general Hamiltonian seems to be difficult. 
It might thus seem that such block-encoding will be especially challenging in our case, as we need to block-encode an entire oscillator (with infinite dimension) into a unitary matrix in order to perform QSP on the oscillator. Surprisingly, some physical interactions between quantum systems can provide natural block-encodings of one system in the basis of the other for qubit-oscillator systems, as is stated formally in Lemma~\ref{def:qubitization-mode}.

\begin{lemma}[Qubitization of a Bosonic Mode via Qubit-oscillator Physical Interaction] Coupling between a qubit Pauli operator $\hat{\sigma}_z$ and a bosonic mode's quadrature operators, $h(\hat{x}, \,\hat{p})$, naturally block-encodes the bosonic mode's unitary evolution operator $\omega(\hat{x}, \,\hat{p}) = e^{-i h(\hat{x}, \,\hat{p}) t}$.
\label{def:qubitization-mode}
\end{lemma}

The above statement immediately follows if we write the resulting unitary under the representation of the qubit's $SU(2)$ matrix, 
\begin{align}
    W_z := e^{-i h(\hat{x}, \,\hat{p}) \hat{\sigma}_z t} 
    = 
    \begin{bmatrix}
        \omega(\hat{x}, \,\hat{p}) & 0\\
        0 & \omega^{-1}(\hat{x}, \,\hat{p})
    \end{bmatrix}.
    \label{wz}
\end{align}
Note that the choice of $\hat{\sigma}_z$ here is only a convention, and any coupling can always be rotated into the $\hat{\sigma}_z$ representation. Also, note that $\omega(\hat{x}, \,\hat{p})$ is an \emph{operator} on the oscillator rather than a complex number. 

Now, given the qubitization of a bosonic mode, we are ready to state a bosonic quantum signal processing theorem that summarizes the achievable polynomial transformations on the quadrature operator $\omega(\hat{x}, \,\hat{p})$, as defined in Lemma~\ref{def:qubitization-mode}.

\begin{theorem}[Bosonic Quantum Signal Processing] 
The following quantum circuit parameterized by $\vec{\theta} = \{\theta_0, \,\ldots, \,\theta_d\}$ achieves a block-encoding of a degree-$d$ Laurent polynomial transformation on $\omega(\hat{x}, \,\hat{p})$ as $F(\omega)$
\begin{align}
    \text{Q}_{\vec{\theta}}(\omega) = e^{i \theta_d \hat{\sigma}_x} \prod_{j = 0}^{d - 1}  W_z e^{i \theta_j \hat{\sigma}_x}  = 
    \begin{bmatrix}
        F(\omega) & i G(\omega) \\
        i G(1 / \omega) & F(1 / \omega)
    \end{bmatrix},
    \label{qsp_theta}
\end{align}
where (setting $t = 1$ for simplicity)
\begin{subequations} \label{eq:fwgw-poly}
    \begin{alignat}{2}
        F(\omega) &= \sum_{n = -d}^d f_n \omega^n = \sum_{n = -d}^d f_n e^{-i h(\hat{x}, \,\hat{p})n} &&:= f(\hat{x}, \,\hat{p}), \label{eq:fw-poly} \\
        G(\omega) &= \sum_{n = -d}^d g_n \omega^n = \sum_{n = -d}^d g_n e^{-i h(\hat{x}, \,\hat{p}) n} &&:= g(\hat{x}, \,\hat{p}). \label{eq:gw-poly}
    \end{alignat}
\end{subequations} 
for $n = \{-d, \,-d + 2, \,-d + 4, \,\ldots, \,d\}$, $f_n, \,g_n \in \mathbb{R}$, $F(\omega) F(1 / \omega) + G(\omega) G(1 / \omega) = 1$, and $h(\hat{x}, \,\hat{p})$ is an analytical function of the bosonic mode's quadrature operators. Inversely, given $F(\omega)$ in Eq.~\eqref{eq:fw-poly} and $F(\omega) F(1 / \omega) < 1$, there exists $\vec{\theta} = \left\{\theta_0, \,\ldots, \,\theta_d\right\}$ such that the construction in Eq.~\eqref{qsp_theta} block-encodes $F(\omega)$.
\label{thm:bosonic-qsp}
\end{theorem}
The proof of Theorem~\ref{thm:bosonic-qsp} follows from the normal QSP proof on single qubit \cite{low2019hamiltonian} or the periodic function formulation \cite{haah_product_2019} once $\text{Q}_{\vec{\theta}}(\omega)$ is expanded under the infinite sets of eigenstates of $h(\hat{x}, \,\hat{p})$. A detailed proof can be found in App.~\ref{app:bosonic-qsp-proof}. Note that a recursive relationship for computing the coefficients $f_n$ and $g_n$ from the phase sequence $\vec{\theta}$ is given in App.~\ref{app:qsp-coeff-recursive}. We also note that despite the similarity to single-qubit QSP, bosonic QSP is formally an infinite-dimensional theorem.

In general, $h(\hat{x}, \,\hat{p})$ can be any physically realizable Hamiltonian of the oscillator (i.e., not only finite degree polynomials but also analytic functions). To the lowest order, $h(\hat{x}, \,\hat{p})$ can be a linear function of $\hat{x}$ and $\hat{p}$, which generates a displacement in the phase space: $ h(\hat{x}, \,\hat{p}) = \alpha \hat{a}^\dagger - \alpha^* \hat{a}$.
Coupling $h$ to a qubit Pauli operator generates a qubit-controlled displacement. Consider the special case where a simple qubit-oscillator coupling naturally arises on cQED hardware \cite{blais2004cavity} or in trapped ions \cite{monroe1996}; we have $h(\hat{x}, \,\hat{p}) t = -\kappa \hat{x}$, where $t$ denotes the duration of the qubit-oscillator interaction or coupling, which generates a displacement operator to boost the oscillator momentum by an amount of $\kappa$. From Lemma~\ref{def:qubitization-mode}, it is readily realized that the physical dynamics generated from this coupling Hamiltonian form a block-encoding of the oscillator operator $\omega := e^{i \kappa \hat{x}}$. Combined with Theorem~\ref{thm:bosonic-qsp}, we have
\begin{subequations} \label{eq:fwgw-poly-short}
    \begin{alignat}{2}
        F(\omega) &= \sum_{n = -d}^d f_n e^{i \kappa \hat{x} n} &&:= f(\hat{x}),  \\
        G(\omega) &= \sum_{n = -d}^d g_n e^{i \kappa \hat{x} n} &&:= g(\hat{x}), 
    \end{alignat}
\end{subequations} 
where $n = \{-d, \,-d + 2, \,-d + 4, \,\ldots, \,d\}$, and $d$ is the degree of QSP, as specified in Theorem~\ref{thm:bosonic-qsp}. The achievable functions and parity constraints upon $\text{Q}_{\vec{\theta}}(\omega)$ for $\omega \in \mathbb{C}$ are described in \cite{haah_product_2019}. Note that in this case $\omega$ is a unitary operator that maps oscillator position $x\in (-\infty, \,\infty)$ to the complex unit circle, so $f(\hat{x})$ and $g(\hat{x})$ are periodic functions with a period $T_x = \frac{2 \pi}{\kappa}$. For an integer $m$:
\begin{subequations} \label{eq:fwgw-periodicity}
    \begin{alignat}{1}
        f(\hat{x} + m T_x) &= f(\hat{x}), \\
        g(\hat{x} + m T_x) &= g(\hat{x}).
    \end{alignat}
\end{subequations}

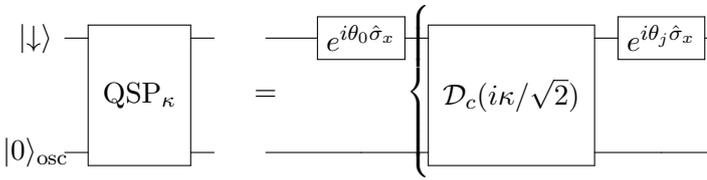
\begin{figure}[htbp]
            \centering
            \mbox{
                \Qcircuit @C=0.8em @R=0.5em @!R {
                    \lvert \downarrow\rangle \quad\quad         & \multigate{2}{\rm QSP_\kappa} & \qw &  & \gate{e^{i \theta_0 \hat{\sigma}_x}} & \multigate{2}{\mathcal{D}_c(i \kappa / \sqrt{2})} & \gate{e^{i \theta_j \hat{\sigma}_x}}   & \qw \\
                    & & &\push{\rule{.3em}{0em}=\rule{.3em}{0em}} & & \\
                    \lvert0\rangle_{\rm osc}\quad\quad & \ghost{\rm QSP_\kappa} & \qw & & \qw & \ghost{\mathcal{D}_c(i \kappa / \sqrt{2})}  & \qw & \qw \gategroup{1}{6}{3}{7}{.7em}{\{} \gategroup{1}{6}{3}{7}{.7em}{\}}
                }
            }
            \caption{A bosonic QSP circuit composed of single-qubit rotations and controlled displacement operations, where the form of $\mathcal{D}_c(i \kappa / \sqrt{2})$ is given in Eq.~\eqref{Dc-def} . The gates inside the bracket are repeated $d$ times for different $\theta_j$ ($j = 1, \,2, \,\ldots, \,d$) in order to obtain a degree-$d$ Laurent polynomial.}
            \label{fig:qsp_circuit}
\end{figure}

The overall construction of the bosonic QSP circuit from Theorem~\ref{thm:bosonic-qsp} is shown in Fig.~\ref{fig:qsp_circuit}, where the conditional displacement operator and single qubit rotations are performed repeatedly. Since this construction performs an arbitrary degree-$d$ real Laurent polynomial on $\omega$ with definite parity, it follows that the resulting functions $f(\hat{x})$ and $g(\hat{x})$ have flexibility to achieve a wide class of functions on $\hat{x}$ in the interval $\left[-\frac{\pi}{\kappa}, \,\frac{\pi}{\kappa}\right]$ that admit at most a degree-$d$ Fourier expansion, as in Eq.~\eqref{eq:fw-poly} and Eq.~\eqref{eq:gw-poly}. Note that the numerical and experimental realization of such conditional displacements in Ref.~\cite{eickbusch2022fast} for universal control of oscillators provides an example of the expressivity of such a bosonic QSP construction. The ability to obtain such nonlinear transformations on oscillator quadrature operators forms the basis of the QSP interferometry, as we will discuss next.

\subsection{QSP Interferometry}
\label{sec:qsp-interferometry}

Building on the ability to perform polynomial transformations on a bosonic oscillator's quadrature operators using QSP in Sec.~\ref{sec:bosonic-qsp} and Theorem~\ref{thm:bosonic-qsp}, we construct a QSP interferometry (QSPI) protocol in this section by combining two bosonic QSP sequences.

A QSP interferometry protocol is defined as follows:
\begin{definition}[Degree-$d$ Quantum Signal Processing Interferometry ($d$-QSPI)] Given an underlying bosonic signal unitary $S_{\beta} = e^{i h_{\beta}(\hat{x}, \,\hat{p})}$, where $h_{\beta}(\hat{x}, \,\hat{p})$ is a finite-degree Hermitian polynomial of the quadrature operators $\hat{x}, \,\hat{p}$ parameterized by $\beta \in \mathbb{R}$, a degree-$d$ quantum signal processing interferometry ($d$-QSPI) protocol for $S_{\beta}$ is defined as the protocol in Fig.~\ref{fig:bosonic-QSP-sensing}c, or $\text{Q}^{-1}_{\vec{\theta}}(\omega) S_{\beta} \text{Q}_{\vec{\theta}}(\omega)$, where $\text{Q}_{\vec{\theta}}(\omega)$ is given by Eq.~\eqref{qsp_theta}. Furthermore, we define the joint qubit-oscillator state created by $\text{Q}_{\vec{\theta}}(\omega)$ as the QSPI sensing state $\ket{\tilde{\text{Q}}} = \text{Q}_{\vec{\theta}}(\omega) \ket{\downarrow}\otimes\ket{0}_{\rm osc}$.
\label{def:qspi}
\end{definition}

It is readily recognized that the QSPI protocol described in Definition~\ref{def:qspi} is a simple generalization of the usual cat-state interferometry protocol, where the cat-state preparation is replaced by an arbitrary bosonic QSP transformation. The QSPI protocol may also be viewed as a parameterized version of typical quantum parameter estimation and discrimination protocols, which involve optimization of a cost function over probe states and measurement operators \cite{helstrom1969quantum}. Each $d$-QSPI protocol is entirely characterized by the angle sequence $\vec{\theta}$, where the first QSP sequence ${\rm QSP}_\kappa$ in Fig.~\ref{fig:bosonic-QSP-sensing}c prepares the optimal sensing state while the second QSP sequence ${\rm QSP}_\kappa^{-1}$ transforms over the measurement probes. Thus, we will minimize our cost function, the probability of decision error, over the QSP angle sequence. In this perspective, our protocol is restricted to be symmetric in the state preparation and measurement procedures (because ${\rm QSP}_\kappa$ and ${\rm QSP}_\kappa^{-1}$ in Fig.~\ref{fig:bosonic-QSP-sensing}c are parameterized by the same set of phase angles), but given the symmetry of the response function we seek this restriction simply makes the optimization more feasible. This choice is also motivated by the success of Ramsey-type protocols, which treat the state preparation and measurement steps symmetrically, but our bosonic QSP framework is able to handle more general approaches taken in other quantum estimation/discrimination schemes by using a different set of QSP angles for the signal decoding operator.

As an interferometer, the unitary operation realized by a $d$-QSPI must be a linear combination of many elementary unitaries, where the interference among them performs some desired quantum sensing task. To make the interferometry aspect of QSPI more vivid and to reveal how individual unitaries (or ``paths'') interfere, it is useful to identify what such elementary unitaries look like in QSPI. Because each matrix element of the bosonic QSP constructed in Eq.~\eqref{qsp_theta} is a linear combination of $\omega^n$ for different $n \in [-d, \,d]$, we define a \emph{QSPI elementary transformation} on the signal $S_{\beta}$ under the basis $\omega^n$:
\begin{definition}[QSPI Elementary Transformation] An elementary QSPI transformation $S_{\beta, \,nm}(\hat{x}, \,\hat{p})$ on $S_{\beta}$ is defined as follows:
\begin{align}
    S_{\beta, \,nm}(\hat{x}, \,\hat{p}) := \omega^{-n} S_{\beta} \omega^{m},
    \label{eq-qspi-et}
\end{align}
for some $\omega = e^{-i h(\hat{x}, \,\hat{p})}$ acting on a bosonic mode and for integers $n, \,m$.
\label{def:qspi-et}
\end{definition}
From a basis-set-expansion point of view, Eq.~\eqref{eq-qspi-et} is equivalent to expanding the unknown operator $S_{\beta}$ under the basis set $\{\omega^n\}$, where $S_{\beta, \,nm}(\hat{x}, \,\hat{p})$ is simply the resulting matrix element (despite its infinite-dimensional nature). In the special case of $\omega = e^{i\kappa\hat{x}}$, Eq.~\eqref{eq-qspi-et} can be viewed as a plane-wave expansion of $S_{\beta}$ for a set of discrete reduced momenta $n k$ for integer $n$, which is similar to the $k$-point sampling technique in the numerical study of periodic solid-state systems \cite{ashcroft2022solid}, where $k$ determines the low-energy cutoff, while the upper limit of $n$ dictates the high-energy cutoff. We will see in the following that this high-energy cutoff is directly related to the QSP degree $d$ that is being used in our construction.

With the elementary transformation defined, we are now ready to present the QSPI transformation theory that describes how QSPI acts as an interferometer:
\begin{theorem}[QSP Interferometry Theorem]
A $d$-QSPI protocol for a bosonic signal unitary $S_{\beta}$ performs a transformation to $S_{\beta}$ such that the resulting unitary is a linear combination of $(d + 1)^2$ elementary transformations $S_{\beta, \,nm}(\hat{x}, \,\hat{p})$ as defined in Def.~\ref{def:qspi-et}
\begin{align}
    \text{Q}^{-1}_{\vec{\theta}}(\omega) S_{\beta} \text{Q}_{\vec{\theta}}(\omega)
    = \sum_{n, \,m = -d}^d C_{nm} S_{\beta, \,nm}(\hat{x}, \,\hat{p}),
    \label{eq-qspit}
\end{align}
where $C_{nm}$ is a complex coefficient matrix defined from the original QSP coefficients by
\begin{align}
    C_{nm} = 
    \begin{bmatrix}
        f_n f_{m} + g_{-n} g_{-m} & i (f_n g_{m} - g_{-n} f_{-m}) \\
        i (f_{-n} g_{-m} - g_n f_{m}) & g_{n} g_{m} f_{-n} f_{-m}
    \end{bmatrix},
    \label{eq-cnm}
\end{align} 
which serves as a complex weight to its associated elementary transformation $S_{\beta, \,nm}$ in order to produce the desired interference.
\label{thm:qspit}
\end{theorem}
The proof of the theorem follows by direct multiplication of the left-hand side of Eq.~\eqref{eq-qspit}. \qed

The right-hand side of Eq.~\eqref{eq-qspit} is a sum of $(d + 1)^2$ terms, each weighted by a complex coefficient matrix $C_{nm}$, as defined in Eq.~\eqref{eq-cnm}; this readily reveals that the resulting unitary of a $d$-QSPI protocol is essentially a giant interferometer of $(d + 1)^2$ elementary components. Note that each element of $C_{nm}$ is always quadratic in terms of $f$ and $g$ (either a product of two $f$ or two $g$ with different subscripts); this is simply a consequence of the fact that the QSP sensing state $\ket{\tilde{\text{Q}}}$ prepared in Fig.~\ref{fig:bosonic-QSP-sensing}c is perturbed by $S_\beta$ before interfering with itself.
In this fashion, the original QSP coefficients $f_n, \,g_n$ can be tuned such that the desired interference pattern is produced by the protocol for any quantum sensing purpose. Note that there is no approximation, such as truncating the dimension of the infinite dimensional oscillators, in our formalism, since we explicitly work with the oscillator quadrature operators, and the physical regularization of the infinite dimensional transformation is provided by a finite-energy initial state (for example, a vacuum state).

Theorem~\ref{thm:qspit} characterizes at the operator level how QSPI works, but it is not clear what or how much information can be extracted from the entire protocol via measurement. We discuss the measurement aspect in the following.

Just as in any interferometry protocol, we are interested in extracting information about $S_{\beta}$ by performing some \emph{measurement} after the protocol, where the measurement outcome contains information about the parameter $\beta$. It is possible to measure oscillators directly using homodyne/heterodyne detection or by performing a photon-number-resolved measurement. Such measurements will often provide \emph{local} information on the phase-space distribution of the oscillator wave function. Alternatively, it is much easier (and typically faster) to measure the ancilla qubit directly. Such a qubit measurement implies a partial trace over the bosonic quadrature operator and therefore will provide useful $\emph{global}$ information about the oscillator. Extracting such global information is crucial to performing decision-making regarding the underlying signal parameter $\beta$, as we will see in Sec.~\ref{sec:disp-sense-bin-dec}. For the ease of discussion, we define the QSPI response function as follows:

\begin{definition}[QSPI Response Function] A QSPI response function is defined as the probability distribution over the signal parameter $\beta$ after a projective measurement on the ancilla qubit as $\mathbb{P}(\beta) = \lVert \bra{\phi_1} \text{Q}^{-1}_{\vec{\theta}}(\omega) S_{\beta} \text{Q}_{\vec{\theta}}(\omega) \ket{\phi_0}\ket{0}_{\rm osc} \rVert^2$, where $\ket{\phi_0}$ and $\ket{\phi_1}$ are the initial and final state of the ancilla qubit and the oscillator is assumed to start from vacuum $\ket{0}_{\rm osc}$.
\label{def:qspir}
\end{definition}

QSPI response functions, or simply the response functions, as defined in Definition~\ref{def:qspir}, characterize the complicated interference pattern between two oscillator states that are perturbed by $S_{\beta}$. The signature of such interference is cast onto the qubit measurement probability. In our case, the unitary channel that we wish to distinguish is a displacement of the oscillator perpendicular to $W_z$'s displacement direction. The effect of the displacing signal $S_\beta$ can be understood as convolving the QSP sensing state $\ket{\tilde{\text{Q}}}$ with a shifted version of itself. Thus, for an optimal choice of the QSP polynomial, we expect to be sensitive to a certain range of displacements. This protocol takes Ramsey interferometry protocols as in \cite{gilmore2021quantum, marciniak_optimal_2022} as inspiration. In fact, Eq.~\eqref{prob-cat-state-sensing} in Sec.~\ref{sec:cat-state-sensing} can be viewed as a simple response function for cat-state sensing, since it can be realized by a degree-1 QSPI protocol, as we will see in the next section.

\subsection{QSPI Protocols for Displacement Sensing}
\label{sec:qspi-displacement}

We will examine the outcome for the qubit state in our generalized QSP sensing protocol for displacement sensing in Sec.~\ref{sec:general-prob}, and use a degree-1 case as an example to connect to the cat state sensing protocol in Sec.~\ref{sec:qspi-degree-1}.

\subsubsection{General Theory of QSPI Displacement Sensing}
\label{sec:general-prob}

We shall drop $\vec{\theta}$ subscripts from $Q$ from here on for simplicity. The sensing sequence, also from Fig.~\ref{fig:bosonic-QSP-sensing}c, is given by:
\begin{align}
     U(\beta, \,\omega) = \text{Q}^{-1}(\omega) S_{\beta} \text{Q}(\omega),
     \label{qsp_conj}
\end{align}
where $\text{Q}^{-1}(\omega) = \text{Q}^{\dagger}(1/\omega)$.
Using Eq.~\eqref{commutation-x-p-exp},
we may rewrite Eq.~\eqref{qsp_conj} as
\begin{align}
    U(\beta, \,\omega) = \text{Q}^{-1}(\omega) S_{\beta} \text{Q}(\omega) = S_{\beta}  \text{Q}^{\dagger}(1/\omega') \text{Q}(\omega),
    \label{eq:q_omega'omega}
\end{align}
where
\begin{align}
    \omega' = \omega e^{-i \kappa \beta} = e^{i \kappa (\hat{x} - \beta)}.
    \label{eq:omega'-def}
\end{align}
The right-hand side of Eq.~\eqref{eq:q_omega'omega} reveals a key insight: the total QSPI protocol reduces to a product of $\text{Q}^{\dagger}(1 / \omega') \text{Q}(\omega)$ (up to an irrelevant global phase $S_{\beta}$), which is a QSP sequence interfering with a shifted version of itself $\omega \rightarrow \omega'$ by a constant phase $e^{-i\kappa \beta}$, as defined in Eq.~\eqref{eq:omega'-def}. It is this $\beta$-dependent shift that allows the extraction of useful information on $\beta$ from the interferometry.

In order to find the response function for the probability of measuring the qubit in the ground state after the sensing protocol, we must integrate over the probability distribution in phase space. Let us denote the upper left matrix element of $U$ as $U_{00} = e^{i \beta \hat{p}}\big[f(-\hat{x} + \beta) f(\hat{x}) + g(\hat{x} - \beta) g(-\hat{x}) \big]$, then the measurement probability of the qubit being at state $\ket{\downarrow}$ is

\begin{align}
    &\mathbb{P}(M = \ \downarrow | \beta) \nonumber \\
    =& \lVert \bra{\downarrow} \text{Q}^{-1} S \text{Q} \ket{\downarrow} \ket{0}_{\rm osc} \rVert^2 \nonumber \\
    =& \langle 0|_{\rm osc} (U_{00})^\dagger U_{00} \ket{0}_{\rm osc} \nonumber \\
    =& \int_{-\infty}^{\infty} dx \lvert\left[f(-x + \beta) f(x) + g(x - \beta) g(-x) \right]\psi_{0}(x) \rvert^2.
    \label{p00_def}
\end{align}
where $\psi_0(x) = \pi^{-1 / 4} e^{-x^2 / 2}$ is the vacuum state of the oscillator, and we have used real numbers $x$ as the argument of $f(\cdot)$ and $g(\cdot)$ since everything has been written under the position representation. 
Thus, $\mathbb{P}(M = \ \downarrow| \beta)$ is a function of our signal parameter $\beta$ and the original QSP phase angles. 
We may now tailor the shape of the QSP Laurent polynomial such that the qubit response $\mathbb{P}(M = \ \downarrow | \beta)$ scales optimally versus $\beta$.

Using the Laurent polynomial expressions from Eq.~\eqref{eq:fw-poly} and Eq.~\eqref{eq:gw-poly} and explicitly evaluating the integration with respect to $x$, we can alternatively write Eq.~\eqref{p00_def} as a series sum
\begin{align}
    \mathbb{P}(M = \ \downarrow | \beta)
    =  \sum_{s = -d}^{d} c_s  \nu(\beta)^s
    \label{p00_series_sum}
\end{align}
for $\nu(\beta) = e^{i (2 \kappa) \beta}$ and $c_s \in \mathbb{R}$ being a function of $\kappa$
\begin{align}
     c_s
    =& \sum_{n, \,n', \,r = -d}^d (f_n f_{n'} + g_n g_{n'}) \nonumber \\
      &\times (f_{n + 2s} f_{n' + 2r} + g_{n + 2s} g_{n' + 2r}) e^{-\kappa^2 (r - s)^2}
    \label{cs-def}
\end{align}
where $n, \,n'$ are either all odd or all even depending on the parity of $d$. It follows that the response function of the qubit $\mathbb{P}(M = \ \downarrow | \beta)$ is a degree-$d$ Laurent polynomial with respect to the new ``signal'' operator $\nu(\beta, \,\kappa) = e^{i (2 \kappa) \beta}$. See Appendix~\ref{app:perr-real-proof} for a proof.

\begin{figure}[htb]
            \includegraphics[width=0.48\textwidth]{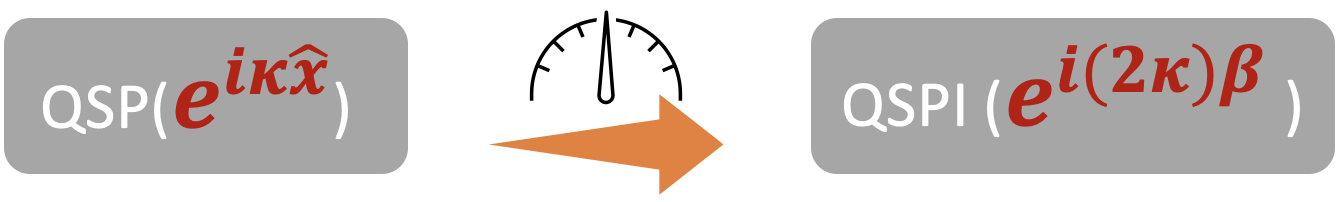}
            \caption{Pictorial illustration of how in the bosonic QSP interferometric protocol, the qubit measurement enacts a duality between a polynomial transformation on the bosonic quadrature operators and a polynomial transformation on the sensing parameter $\beta$ via QSPI.
            }
            \label{fig:duality-cartoon}
\end{figure}

Therefore, apart from parity and normalization constraints, we may design the qubit response $\mathbb{P}\left(M = \ \downarrow | \beta\right)$ by choosing $f_n$ and $g_n$ such that we approach the desired Fourier series of a function of $\beta$. This relationship also reveals an interesting duality between the QSP polynomial transformation on phase space quadrature and the polynomial transformation on the signal parameter $\beta$ in the response function, which is highlighted in Fig.~\ref{fig:duality-cartoon} and summarized as Theorem~\ref{thm:qsp-displacement-sensing}.

\begin{theorem}[QSPI for Displacement Sensing]
Given a degree-$d$ QSPI protocol with the block-encoded quadrature operator $\hat{\omega} = e^{i \kappa \hat{x}}$ (periodic with a period of $\left[-\frac{\pi}{\kappa}, \,\frac{\pi}{\kappa}\right]$ with respect to $x$), a degree-$d$ response function $\mathbb{P}(\nu) := \sum_s c_s \nu^s$, with $\nu = e^{i (2 \kappa) \beta}$ as its argument, is well-defined where $\beta \in \left[-\frac{\pi}{2\kappa}, \,\frac{\pi}{2 \kappa}\right]$. Conversely, given a degree-$d$ real Laurent polynomial transformation  (defined in Eq.~\eqref{p00_series_sum} with $c_s$ as its coefficients) on $\nu = e^{i (2 \kappa) \beta}$ where $\beta \in \left[-\frac{\pi}{2\kappa}, \,\frac{\pi}{2\kappa}\right]$ as the desired response function that satisfies the following necessary conditions, 
\begin{subequations}
    \begin{alignat}{1}
        &\sum_{s = -d}^d c_s = 1, 
            ~~~c_s = c_{-s}, \label{qspir_constraints} \\
        &0 \leq \sum_s c_s \nu(\beta)^s \leq 1,  \label{qspir_constraints-2}
    \end{alignat}
\end{subequations}
there exists a $d$-QSPI protocol as in Fig.~\ref{fig:bosonic-QSP-sensing}c that realizes the desired response function.
\label{thm:qsp-displacement-sensing}
\end{theorem}
We will now sketch a proof of the above theorem. From \cite{haah_product_2019}, we know for a given QSPI protocol characterized by $\vec{\theta}$ that $f(x)$ and $g(x)$ are well-defined, and thus the forward direction of Theorem~\ref{thm:qsp-displacement-sensing} is trivial. 

On the other hand, it is in general challenging to provide sufficient conditions for response functions such that they can be realized by the QSPI protocol using a set of phase angles $\vec{\theta}$. Here for the reverse direction of Theorem~\ref{thm:qsp-displacement-sensing}, we resort to only necessary conditions on the response function. 

Because the response function is necessarily a probability on the qubit and involves a integral over the bosonic coordinates, there are additional constraints on the response function. First, from the response function's relevant parity and normalization constraints on $c_s$ as defined in Eq.~\eqref{p00_series_sum}, Eq.~\eqref{qspir_constraints} can be derived.
Secondly, beyond the parity and normalization constraints, we must also impose an additional constraint on possible sets of $c_s$ due to the response function being a probability, which requires Eq.~\eqref{qspir_constraints-2} to be satisfied for all $\beta \in \mathbb{R}$.
\qed

Let us make additional remarks on the reverse direction of Theorem~\ref{thm:qsp-displacement-sensing}. The reason that it is challenging to ascertain a set of sufficient conditions on the response function is due to the nonlinear transformation of $F, \,G$ in Eq.~\eqref{p00_def}. Inverting the system of polynomial equations in Eq.~\eqref{cs-def}, even in the large $\kappa$ limit where decay coefficients vanish, appears analytically intractable. From our construction in Eq.~\eqref{eq:fwgw-poly}, it can be inductively shown that $f_{-d} = g_{d} = 0$, and from \cite{haah_product_2019}, $F(\omega)$ determines $G(\omega)$ up to $\omega \mapsto 1/\omega$. Thus, for fixed $\kappa$, we may reduce Eq.~\eqref{cs-def} to a system of $d$ independent equations with $d$ unknowns, but searching for solution sets of such polynomial equations is difficult and often infeasible analytically.

\subsubsection{Example: Cat State Sensing}
\label{sec:qspi-degree-1}
For the degree $d = 1$ case, and $\left\{\theta_0, \,\theta_1\right\}$ as the QSP phase angles, the QSPI response function is
\begin{align}
    \mathbb{P}\left(M = \ \downarrow | \beta\right) 
    = c_0 + c_1 \nu + c_{-1} \nu^{-1} 
\end{align}
where 
\begin{align}
    c_0 = \cos^4(\theta_0) + \sin^4(\theta_0) \\
    c_1 = c_{-1} = \cos^2(\theta_0) \sin^2(\theta_0).
\end{align}
Evidently, this is a degree-1 Laurent polynomial of the argument $\nu = e^{i (2 \kappa) \beta}$, where all of the polynomial coefficients are real. Also, we note that the measurement probability is independent of $\theta_1$ (or, in general, $\theta_d$ for a $d$-QSPI protocol) per the construction. Furthermore, by choosing $\theta_0 = \pi / 4$, we recover exactly the cat state sensing protocol with $\mathbb{P}\left(M = \ \downarrow | \beta\right) = \cos^2(\kappa \beta)$. Thus, the cat state sensing protocol is indeed a special case of a 1-QSPI protocol!

We can also view the cat-state sensing protocol for a displacement of $\kappa d$ as the trivial case of choosing all zero rotation angles for $R_X$ in a $d$-QSPI protocol. In this case, the average photon number in the prepared state after the state-preparation unitary in Fig.~\ref{fig:bosonic-QSP-sensing}a will be $n_{\rm photon} \propto \kappa^2 d^2$. Therefore, Def.~\ref{def:hl-decision} reduces to $p_{\rm err} \sim O(1 / \sqrt{n_{\rm photon}})$. This agrees with the results in Sec.~\ref{sec:cat-state-sensing} and Fig.~\ref{fig:d1_compare} (vide infra) that cat states are not efficient resource states for the Main Problem.

\section{Binary Decision-Making Using QSPI}
\label{sec:thres-binary-decision}

Now, we have established the QSPI protocol and its response function as a polynomial transformation of the signal. To demonstrate its potential for general sensing tasks, we derive an explicit expression for the response function for a binary decision problem on the displacement parameter in Sec.~\ref{sec:disp-sense-bin-dec} and analyze analytically the decision quality and sensing complexity in Sec.~\ref{sec:binary-decision-making}.

\subsection{Binary Decision for Displacement Sensing}
\label{sec:disp-sense-bin-dec}

For a classical binary decision using measurement of a single qubit, we want the QSPI response function to be either 1 or 0 depending on the value of the signal displacement relative to $\beta_{\rm th}$.  In particular, one such target qubit response function is the step function
    \begin{equation}
        P_{\rm ideal}(\beta)  =
        \begin{cases} 
            1, & 0 \le \lvert\beta\rvert < \beta_{\rm th} \\
            0, & \beta_{\rm th} < \lvert\beta\rvert \le \frac{\pi}{2 \kappa}.
        \end{cases}
        \label{eq:pideal}
   \end{equation}
If such an ideal qubit response function is realized, then the binary decision sensing protocol has no error. However, in practice, only a finite-degree polynomial approximation to this function is available and lead to decision errors. In the following, we give basic definitions to quantify the decision errors.

\begin{figure}[htbp]
            \includegraphics[width=0.48\textwidth]{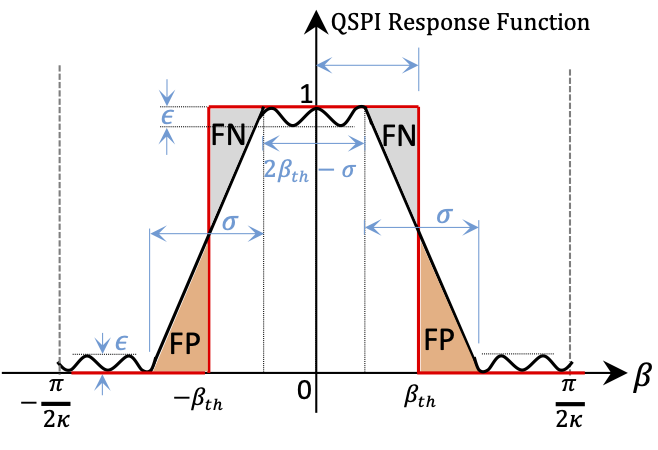}
            \caption{Schematic of erroneous decision making probability (from the response function) as the difference between the ideal response function (solid red) and a polynomial approximated response function (solid black). For an event defined as ``the displacement is below threshold $\beta_{\rm th}$,'' the integrated erroneous probability is composed of two parts: i) missing the event while it actually happened (false-negative, FN, Type-II error, grey-shaded area), ii) reporting the event when it did not happen (false-positive, FP, Type-I error, orange-shaded area). Note that the effective detection signal range is $\left[-\frac{\pi}{2\kappa}, \,\frac{\pi}{2\kappa}\right]$.}
            \label{fig:perr-schematic}
\end{figure}

For ease of discussion, Fig.~\ref{fig:perr-schematic} plots the ideal response function (red) in contrast to a typical polynomial approximation generated by the QSPI protocol (black) as a function of the underlying displacement $\beta$. The approximated response function features a steep yet finite slope centered about $\beta_{\rm th}$. There are usually some small oscillatory patterns for small $\beta$ and for $\beta_{\rm th} < \lvert \beta \rvert < \frac{\pi}{2\kappa}$. Deviations of $P_{\rm approx}(\beta)$ from $P_{\rm ideal}(\beta)$ across the entire range of $\left[-\frac{\pi}{2 \kappa}, \,\frac{\pi}{2 \kappa}\right]$ quantify the overall probability of making the wrong decision. We define the following quantity as \emph{decision error density}:
\begin{align}
    p_{\rm err}\left(\beta_{\rm th}, \,\kappa\right) 
    &= \frac{\kappa}{\pi} \int_{-\frac{\pi}{2 \kappa}}^{\frac{\pi}{2 \kappa}} 
     \lvert P_{\rm approx}(\beta) - P_{\rm ideal}(\beta) \rvert d\beta \nonumber \\
    &= p_{\rm err, \,FN}(\beta_{\rm th}) + p_{\rm err, \,FP}(\beta_{\rm th}).
    \label{perr-def}
\end{align}
This quantity can be split into two contributions according to $\beta_{\rm th}$ as follows:
\begin{subequations}\label{eq:perr-def-miss-detection-false-alarm}
    \begin{alignat}{1}
        p_{\rm err, \,FN}(\beta_{\rm th}) &= \frac{2\kappa}{\pi} \int_{0}^{\beta_{\rm th}} \left(1 - P_{\rm approx}(\beta) \right) d\beta, \label{eq:perr-def-miss-detection} \\
        p_{\rm err, \,FP}(\beta_{\rm th}) &= \frac{2\kappa}{\pi} \int_{\beta_{\rm th}}^{\pi/2\kappa} P_{\rm approx}(\beta)  d\beta. \label{eq:perr-def-false-alarm}
    \end{alignat}
\end{subequations}
The former, $p_{\rm err, \,FN}(\beta_{\rm th})$, defined in Eq.~\eqref{eq:perr-def-miss-detection}, is the false-negative (FN) error (also called Type-II error in hypothesis testing) and is indicated by the grey region in Fig.~\ref{fig:perr-schematic}, while the latter, $p_{\rm err, \,FP}(\beta_{\rm th})$, defined in Eq.~\eqref{eq:perr-def-false-alarm}, is the false-positive (FP) error (Type-I error) and is highlighted in orange in Fig.~\ref{fig:perr-schematic} \cite{helstrom1969quantum}. Our goal in designing the response function is to find QSP phase angles that minimize the total decision error in Eq.~\eqref{perr-def}.

Substituting Eq.~\eqref{p00_series_sum} into Eq.~\eqref{perr-def}, the error probability can be written explicitly as
\begin{align}
    p_{\rm err}(\beta_{\rm th}, \,\kappa) = \sum_{s = -d}^{d} c_s H_s(\beta_{\rm th}, \,\kappa),
    \label{perr-analytical}
\end{align}
where we have defined
\begin{align}
    H_s(\beta_{\rm th}, \,\kappa) = \frac{2 \kappa \beta_{\rm th}}{\pi} + \sinc\left(\pi s \right) - \frac{4 \kappa \beta_{\rm th}}{\pi} \sinc\left(2 \kappa s \beta_{\rm th} \right)
    \label{H_s_expression}
\end{align}

\noindent as a function of $(\beta_{\rm th}, \,\kappa)$, and the definition of $c_s$ is given in Eq.~\eqref{cs-def}. Eq.~\eqref{perr-analytical} is the central metric for the binary decision problem in displacement sensing.

Additionally, note that $\kappa$ and $\beta_{\rm th}$ always appear together in Eq.~\eqref{H_s_expression} as a product, which suggests there is a scale-invariance in the definition of $H_s$ in the sense that the displacement length scale can be measured in units of $1/\kappa$ and only this relative length scale with respect to $1/\kappa$ is meaningful. Toward this end, we introduce a dimensionless parameter $\eta$ to quantify $\beta_{\rm th} = \eta \frac{\pi}{\kappa}$. It is evident that the dynamic range for the signal $\beta$ will be $\eta \in \left[-0.5, \,0.5\right]$, as is suggested by the horizontal axis of Fig.~\ref{fig:perr-schematic}. However, the dependence of the coefficients $c_s$ on $\kappa$ in Eq.~\eqref{cs-def-app} means $p_{\rm err}$ is still $\kappa$-dependent, which in turn suggests the optimal sensing QSPI phase angles depend on $\kappa$.

Lastly, note that \emph{some} information is inevitably lost when in the final step of the sensing protocol we trace out the oscillator part of the system with the qubit measurement, leaving only the qubit part of the system.  However, by selecting the QSPI phases so as to engineer the polynomials $F$ and $G$ in $\omega$ (Eq.~\ref{eq:fw-poly} and Eq.~\ref{eq:gw-poly}) prior to measurement, we are able to maximize the proximity of the response-function polynomial in $\beta$ after measurement to the ideal step-function response with transition at $\beta_{\rm th}$ (and hence minimize $p_{\rm err}$).  In this way, we ensure that the \emph{relevant} global displacement information is transferred from the oscillator to the qubit, allowing for the efficient extraction of a decision about the displacement magnitude without the need to directly read out the oscillator's state (which would require many samples and measurements).

\subsubsection{Limitation of Cat State for Decision Making}
\label{sec:cat-state-limitation-dec}
As we have seen in Sec.~\ref{sec:qspi-degree-1} and in Fig.~\ref{fig:bosonic-QSP-sensing}b, the cat-state sensing protocol corresponds to a degree $d = 1$ QSPI. It has also been shown in Eq.~\eqref{cat-state-delta-beta} in Sec.~\ref{sec:cat-state-sensing} that the cat state sensing protocol achieves the celebrated Heisenberg-limited sensing for parameter estimation. In this section, we will characterize the performance of cat-state sensing protocol for \emph{decision-making} and reveal its limitations in this regard.

For degree $d = 1$ QSPI, the integrated probability of making the wrong decision per unit signal can be calculated as 
\begin{align}
    p_{\rm err} &= \frac{1}{4 \pi } \big[ \sin (2 \beta_{\rm th} \kappa - 4 \theta_0)+\sin (2 \beta_{\rm th} \kappa + 4 \theta_0) \nonumber \\
            &+ (\pi - 4 \beta_{\rm th} \kappa) \cos (4 \theta_0) - 4 \beta_{\rm th}
   \kappa - 2 \sin (2 \beta_{\rm th} \kappa) + 3 \pi  \big].
\end{align}
We would like to minimize $p_{\rm err}$ overall by choosing $\theta_0$ appropriately. The global minimum is found to be the following when $\theta_0 = \pi / 4$ (regardless of $\beta_{\rm th}$ and $\kappa$):
\begin{align}
    p_{\rm err}|_{\theta_0 = \frac{\pi}{4}} = \frac{1}{2} - \frac{\sin(2 \kappa \beta_{\rm th})}{\pi}. 
\end{align}

On the other hand, supposing that we perform no rotation on the qubit, or $\theta_0 = 0$, we obtain $\mathbb{P}(M = \ \downarrow | \beta) = 1$, which gives 
\begin{align}
    p_{\rm err}|_{\theta_0 = 0} = 1 - \frac{2 \kappa \beta_{\rm th}}{\pi} .
\end{align}
This makes sense because this scenario is equivalent to making a decision that the displacement is always below $\beta_{\rm th}$, and therefore there is only false positive error and the error probability should decrease as $\beta_{\rm th}$ is increased. Moreover, when $\beta_{\rm th} = \frac{\pi}{2 \kappa}$, which is on the boundary of the sensing range, $p_{\rm err}$ drops to zero.

\begin{figure}[htbp]
            \includegraphics[width=0.48\textwidth]{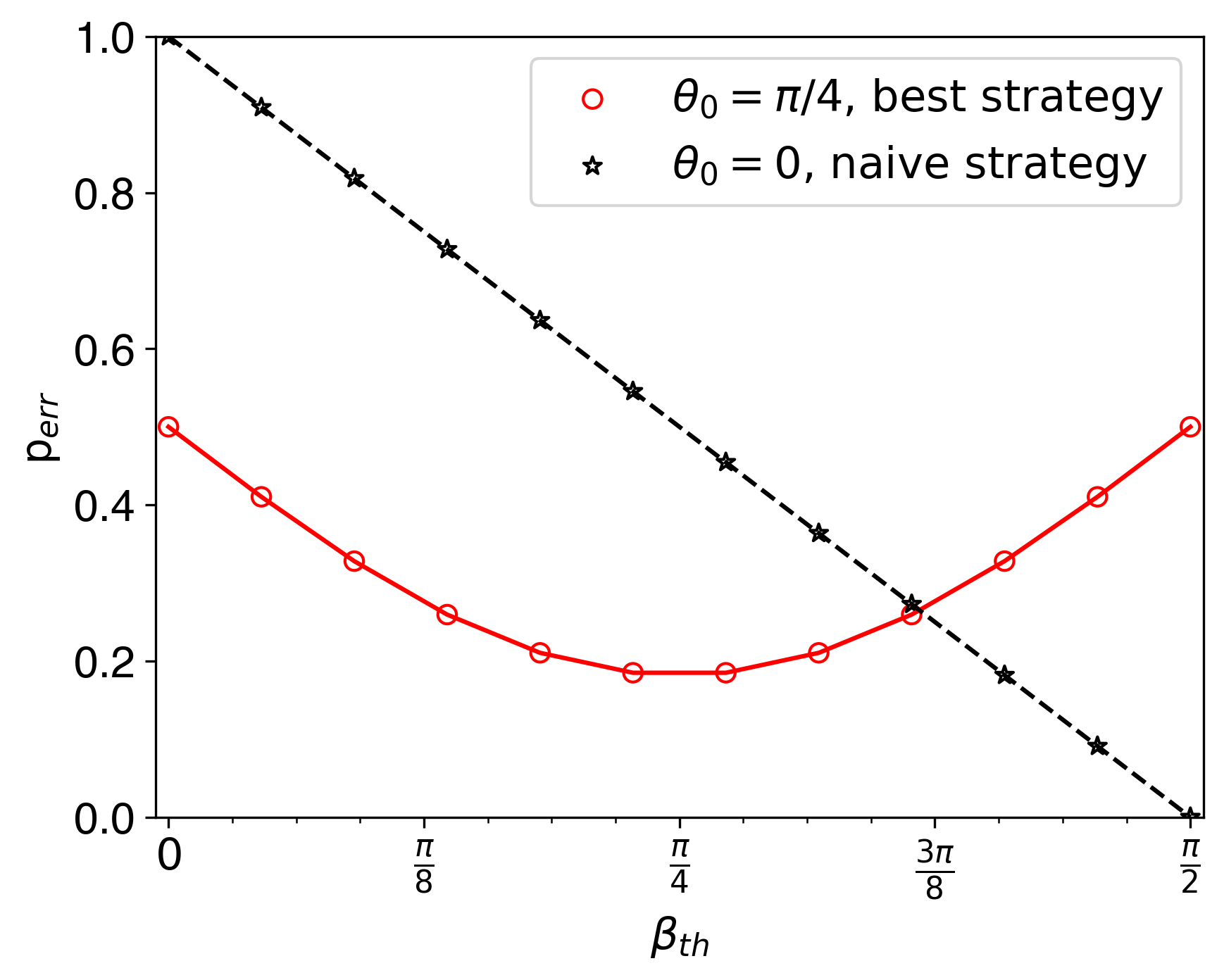}
            \caption{The probability of making wrong decision versus the decision threshold $\beta_{\rm th}$. Data shown for a binary decision of displacement sensing using degree-1 bosonic QSP with $\kappa = 1$, comparing the best ($\theta_0 = \pi / 4$, red circle) and the na\"{i}ve ($\theta_0 = 0$, black star) sensing protocol.}
            \label{fig:d1_compare}
\end{figure}

A comparison of the scaling of $p_{\rm err}$ versus $\beta_{\rm th}$ between the best decision $\theta_0 = \pi/4$ and the ignorant decision $\theta_0 = 0$ is shown in Fig.~\ref{fig:d1_compare}. It can be seen that the optimal sensing strategy significantly reduces $p_{\rm err}$ when $\beta_{\rm th} < \beta_{\rm th}^*$, while it performs worse than the na\"ive guess for $\beta_{\rm th} > \beta_{\rm th}^*$ where $\beta_{\rm th}^* = \frac{\chi}{2 \kappa}$ for $\chi$ being the solution to the transcendental equation $\frac{\pi}{2} - x + \sin(x) = 0$. The optimal sensing protocol works best for $\beta_{\rm th} = \frac{\pi}{4 \kappa}$, where it gives $p_{\rm err} = \frac{1}{2} - \frac{1}{\pi}$. This simple example also demonstrates the non-trivial complexity of decision problems, even for the simplest such problems.

The above analysis also reveals that the minimum $p_{\rm err}$ achieved by the cat-state sensing protocol is always a constant regardless of the value of the displacement $\kappa$ or the size of the cat state for a threshold $\beta_{\rm th} = \frac{\pi}{4 \kappa}$. This behavior is in drastic contrast with parameter estimation tasks, where a larger cat state will have finer interference fringes in phase space and therefore achieve Heisenberg-limited estimation accuracy for very small displacements (Eq.~\eqref{cat-state-delta-beta} of Sec.~\ref{sec:cat-state-sensing}). An intuitive reason why decision-making with a larger cat-state does not help is that the integration in Eq.~\eqref{perr-def} smears out the local information in the oscillator wave function (e.g., the fine interference fringes), meaning that globally, a larger cat-state behaves the same as a smaller cat-state for the binary decision.

\subsection{Algorithmic Complexity for Binary Decision}
\label{sec:binary-decision-making}

Given the definition of the decision error density $p_{\rm err}(\beta_{\rm th}, \,\kappa)$, in this section, we would like to understand some fundamental limits on how $p_{\rm err}(\beta_{\rm th})$ scales with the degree of the QSP. This determines the algorithmic complexity of making a high-quality binary decision using a QSPI protocol.

To do this, we first construct a composite function that exactly reproduces $P_{\rm ideal}(\beta)$ (red trace in Fig.~\ref{fig:perr-schematic}) in the relevant sensing range of $\left[-\frac{\pi}{2\kappa}, \,\frac{\pi}{2\kappa}\right]$.  In particular, consider
\begin{align}
    &P_{\rm ideal}^{\rm sign, \,sin}(\beta) \nonumber \\
    =& \frac{\sign\left[ \sin\left(\kappa\left(\beta_{\rm th} - \beta\right)\right) \right] + \sign\left[ \sin\left(\kappa\left(\beta_{\rm th} + \beta\right)\right) \right]}{2},
\end{align}
where $\sign(\cdot)$ is the sign function. We use the superscript ${}^{\rm sign, \,sin}$ to distinguish the current construction from other possible constructions for $P_{\rm ideal}(\beta)$. 

With this, it can be shown (Appendix~\ref{app:heisenberg-scaling-proof}) that to achieve a target faulty-decision probability $p_{\rm err}$, the required QSP polynomial degree $d$ to approximate $P_{\rm ideal}^{\rm sign, \,sin}(\beta)$ must have
\begin{align}
    d \propto \frac{1}{\kappa p_{\rm err}} \log\left( \frac{1}{\kappa p_{\rm err}}\right).
    \label{d_vs_perr}
\end{align}
For small $p_{\rm err}$, $\log(\frac{1}{p_{\rm err}}) \ll \frac{1}{p_{\rm err}}$. Therefore, the total faulty-decision probability can be solved from Eq.~\eqref{d_vs_perr} 
\begin{align}
    p_{\rm err} \propto \frac{1}{\kappa d} \log(d).
    \label{perr_vs_d}
\end{align}
Recall that in a standard  parameter estimation task, Heisenberg-limited scaling is defined when the standard deviation for estimating the underlying parameter scales as $1/t$ where $t$ is the total time for the sensing protocol (also see discussions in Sec.~\ref{sec:cat-state-sensing}). Here for our case of binary decision making, analogous to parameter estimation, Eq.~\eqref{perr_vs_d} suggests that our QSPI protocol can achieve the efficient scaling defined in Definition~\ref{def:hl-decision}, where the probability of making the wrong decision decreases as $1/d$ (up to a logarithmic factor of $\log(d)$) and $d$ is proportional to the runtime of the sensing protocol. More strictly speaking, the appearance of the $\log(d)$ factor in Eq.~\eqref{perr_vs_d} will make the actual scaling slightly worse than the efficient scaling from Definition~\ref{def:hl-decision}, as is corroborated by numerical evidence in Sec.~\ref{sec:numerical}.

\section{Parameter Estimation from Quantum Decisions}
\label{sec:quantum_decisions}

Among general quantum sensing tasks in addition to decision making, another important class of problems is to estimate a parameter (or multiple parameters) up to a given precision. In Sec.~\ref{sec:thres-binary-decision}, it has been shown that efficient scaling can be achieved in determining whether a displacement signal is below or above a given threshold in the single-sample limit. One natural question is whether such a decision making protocol can be combined with classical search or majority voting techniques to efficiently estimate a given displacement parameter $\beta$ to high precision $\epsilon$. This approach to parameter estimation is akin to \emph{probably approximately correct} metrology in \cite{meyer2023quantum} where they reframe finite-sample quantum metrology in terms of probability that the estimated parameter is within a certain error tolerance of the true value. We approach this problem from the Bayesian perspective and assume a uniform prior distribution for $\beta$. Our $p_{\rm err}$ function is equivalent to the definition of Bayesian success probability $\eta$ in \cite{meyer2023quantum} for uniform prior distribution over parameter $\beta$ in a range $\pi / (2 \kappa)$. 

In this section, we show that this is possible by repeatedly querying the binary decision protocol in Sec.~\ref{sec:disp-sense-bin-dec}. In particular, Sec.~\ref{sec:naive-protocol} presents a na\"{i}ve classical binary search strategy with a finite success probability. Sec.~\ref{sec:majority-vote} then describes a way to exponentially suppress the failure probability using majority voting.

\subsection{Na\"{i}ve Classical Binary Search Protocol}
\label{sec:naive-protocol}

Suppose that we have a displacement channel that results in displacement by some unknown $\beta \in \left[0, \,R\right]$, and we would like to know the value of $\beta$ within error $\delta$ using the quantum decision protocol in previous sections. How long (in terms of QSP degree and queries to the displacement signal) will it take? 

We may approach this task by a na\"{i}ve binary search method, which subdivides the search space at each step by performing a binary decision using an approximate threshold function, as in Sec.~\ref{sec:disp-sense-bin-dec}, to determine whether $\beta$ is above or below the threshold. This process may continue with a different threshold each time until finally we pin down the value of $\beta$ to within $\delta$. Obviously, if a perfect threshold function (red trace in Fig.~\ref{fig:perr-schematic}) is provided each time, this reduces to a purely classical search problem, and it is known that for $N$ total bisections, we can determine $\beta$ to a precision of $\delta = R / 2^N$. Therefore, to achieve fixed accuracy $\delta$ on $\beta$, the required number of queries to the perfect threshold function will be 
\begin{align}
    N_{\rm query, \,ideal} = \log_2(R / \delta),
    \label{n-query-ideal-binary-search}
\end{align}
which demonstrates exponentially increased accuracy as the number of queries to the perfect threshold function increases. The entire protocol will be deterministic with unity success probability.

However in practice, due to the polynomial approximation to the ideal threshold function, there will be a non-zero probability of making the wrong decision at each step. Therefore, the overall binary search will have a finite failure probability. Moreover, the finite width in the falling edge of the qubit response function will necessarily impose a fundamental limit on how accurately the binary search can continue for given fixed QSP degree $d$. 

To perform the binary search, the first step is to choose the relevant $\kappa$ value for the controlled-displacement operator in QSP to set the upper bound on the period of the response function to at least $R$. Thus, we choose $\kappa = \pi/(2 R)$. The position of the threshold can be chosen for the $j$-th binary search to be $\beta_{{\rm th}, \,j} = R / 2^j$. Recall that the width of the falling edge of the qubit response function is $\sigma \approx p_{\rm err}$, which means we are limited to an accuracy of $\delta \approx \sigma \approx p_{\rm err}$ before subsequent thresholds lie within the falling edge of the current qubit response function (i.e., the decision protocol becomes a parameter estimation problem).

Combining Eq.~\eqref{n-query-ideal-binary-search} and Eq.~\eqref{perr_vs_d} provides a limit on the total number of binary decision searches one may perform,
\begin{align}
    N_{\rm query, \,max} \approx \log_2(\kappa d R / \log(d)),
    \label{N-query-max}
\end{align}
before the falling edge's nonzero width prevents us from making a sharp decision.

Moreover, since each binary search has a failure probability of $p_{\rm err}$, the overall failure probability of the binary search protocol for a given degree-$d$ will, for \emph{small} error probability $p_{\rm err}$, be approximately given by
\begin{align}
    p_{\rm failure} 
    = p_{\rm err}  N_{\rm query, \,max}
    \approx \frac{\log(d)}{\kappa d} \log_2\left(\frac{\kappa d R}{\log(d)}\right).
    \label{p-failure}
\end{align}

Therefore, a $d$-QSPI sensing protocol can at best estimate the displacement $\beta$ to precision $\delta = \frac{1}{\kappa d} \log(d)$ with $N_{\rm query, \,max}$ queries to the displacement signal and a limited success probability of $1 - p_{\rm failure}$, where $N_{\rm query, \,max}$ and $p_{\rm failure}$ are given by Eq.~\eqref{N-query-max} and Eq.~\eqref{p-failure}, respectively. Following this procedure, one can of course make further queries to reduce the uncertainty of the underlying parameter $\beta$, but this is more akin to a parameter-estimation protocol rather than a decision protocol.

\subsection{Boosting Estimation Success Probability via Majority Voting}
\label{sec:majority-vote}

The above search protocol has a failure probability that decreases $\sim 1 / d$ as the QSP degree $d$ increases. In this section, we show that the failure probability can be suppressed exponentially as the number of samples increases.

Consider modification of the above protocol by repeating the binary decision $M$ times during each round of search, and then using majority vote to determine the decision outcome in each search. From binomial distribution, the total probability of making the wrong decision for each round of the binary search is
\begin{align}
    p_{{\rm err}, \,M} 
    &\propto \sum_{m = M/2 + 1}^{M} \binom{M}{m} p_{\rm err}^m (1 - p_{\rm err})^{M - m} \nonumber \\ 
    &\approx \frac{M!}{(M / 2)! (M / 2)!} p_{\rm err}^{M / 2}
    \label{perr_mv}
\end{align}
for small $p_{\rm err}$ (or large $d$). Following similar analysis, the overall failure probability of the entire binary search will be reduced to
\begin{align}
    p_{{\rm failure}, \,M}
    &= p_{{\rm err}, \,M}  N_{\rm query, \,max} \nonumber \\
    &\approx \left( \frac{4 \log(d)}{\kappa d} 
    \right)^{M / 2}  \log_2\left(\frac{\kappa d R}{\log(d)}\right) \nonumber \\
    &= O(d^{-M / 2})
    \label{p-failure-M}
\end{align}
which is an exponential suppression as compared to the $1/d$ scaling in the case of single-sample decision, at the expense of a total of $N_{\rm query, max} M$ queries to the displacement signal.

\section{Numerical Results and Discussions}
\label{sec:numerical}

Building on the fundamental theory and analytical analysis of the quantum signal processing interferometry, in this section we provide numerical evidence to support our analytical findings for binary decision-making on displacement channels and to demonstrate the advantage of the QSPI protocol for quantum sensing tasks. 

A key task to construct the desired QSPI protocol is to find the corresponding QSPI phase angles $\vec{\theta}$ as in Definition~\ref{def:qspi}. For a given decision problem, the existing analytic angle-finding algorithms for QSP \cite{geronimo2004positive,chao_finding_2020} cannot be directly applied in QSPI to realize the optimal response function due to reasons mentioned in Sec.~\ref{sec:general-prob}. In this section, we resort to numerical optimization algorithms,
which are capable of carrying out such multi-variable approximate optimization tasks on a reasonable timescale to find the QSPI angles. See Ref.~\cite{qspi2023} for the source code and related data accompanying this work. The rest of this section is organized as follows. Sec.~\ref{sec:numerical-qspir-phases-binary-decision} presents the response function from our numerical optimization, while Sec.~\ref{sec:numerical-heisenberg-scaling} further exhibits the favorable efficient scaling for the error decision probability from these numerical solutions. In Sec.~\ref{sec:wigner-function}, we discuss features of the Wigner function of the optimal sensing states $\ket{\tilde{\text{Q}}}$ as defined in Definition~\ref{def:qspi}.

\subsection{QSPI Phases for Binary Decisions}
\label{sec:numerical-qspir-phases-binary-decision}

    As discussed, we seek a QSPI protocol
    that generates a response function approximating a step function with sharp transitions at $\pm \beta_{\rm th}$, as
    given in Eq.~\eqref{eq:pideal}. 
    We approximate the ideal response function via machine optimization of the phases $\theta_j$ 
    in Eq.~\eqref{qsp_theta} to minimize the objective function $p_{\rm err}$ in Eq.~\eqref{perr-def} and Eq.~\eqref{perr-analytical} for different degrees $d$. We use the standard Nelder-Mead optimization algorithm as implemented in Python in the \texttt{scipy} optimization package with convergence defined to take place whenever either $p_{\rm err}$ or every QSPI rotation angle in $\vec{\theta}$ changes by at most $10^{-5}$ radians in a single step of optimization.

    Using this learned QSPI phase sequence, we can now define an explicit experimental protocol for sensing a displacement as follows:
    \begin{protocol}{QSPI Sensing \label{pro:qspi-sensing}}
                \vspace{-8pt}
        \begin{enumerate}
                    \vspace{-8pt}
            \item \textbf{Parameter selection:} Given $\beta_{\rm th}$ and a range of the signal $\beta \in \left(0, \,\beta_{\rm max}\right)$, pick $\kappa = \frac{\pi}{2 \beta_{\rm max}}$ such that $\beta_{\rm th}$ and $\beta$ fall in the first period of the effective sensing range $\left[0, \,\frac{\pi}{2 \kappa}\right)$ (Fig.~\ref{fig:perr-schematic}). In our discussion, we have picked $\kappa = \frac{\pi}{4 \beta_{\rm th}}$, assuming that $\beta_{\rm max} = 2 \beta_{\rm th}$.
                        \vspace{-8pt}
            \item \textbf{Numerical optimization/phase learning on classical computers:} Perform classical optimization for the desired QSPI phases given this $\beta_{\rm th}$ and $\kappa$ using the code in the QSPI repository \cite{qspi2023}; denote the output phase angles of the code as $\vec{\theta}$.
            \vspace{-8pt}
            \item \textbf{Experimental realization on quantum systems:} Using the experimental implementations of qubit rotations and controlled displacements by $\kappa$, implement the experimental sensing protocol corresponding to the learned QSPI phases $\vec{\theta}$:
            \begin{enumerate}
                        \vspace{-6pt}
                \item \textbf{QSPI State Preparation:} Prepare the QSPI sensing state according to Eq.~\eqref{qsp_theta} and Fig.~\ref{fig:qsp_circuit}.
                \begin{enumerate}
                    \item Perform a qubit rotation by $\theta_0$.
                    \item For $1 \leq j \leq d$:
                    \begin{enumerate}
                        \item Perform a controlled displacement by $\kappa$.
                        \item Perform a qubit rotation by $\theta_j$.
                    \end{enumerate}
                \end{enumerate}
                 \vspace{-6pt}
                \item \textbf{Signal:} Apply the unknown displacement signal for sensing to the oscillator in the qubit-oscillator system.
                 \vspace{-6pt}
                \item \textbf{QSPI Signal Decoding:} Undo the preparation of the QSPI sensing state for readout.
                \begin{enumerate}
                    \item For $d \geq j \geq 1$:
                    \begin{enumerate}
                        \item Perform a qubit rotation by $-\theta_j$.
                        \item Perform a controlled displacement by $-\kappa$.
                    \end{enumerate}
                    \item Perform a qubit rotation by $-\theta_0$.
                \end{enumerate}
                \item \textbf{QSPI Readout:} Measure the ancilla qubit under the Pauli-$Z$ basis once.
            \end{enumerate}
             \vspace{-8pt}
        \end{enumerate}
    \end{protocol}
    
\begin{figure}[htbp]
        \includegraphics[width=\columnwidth]{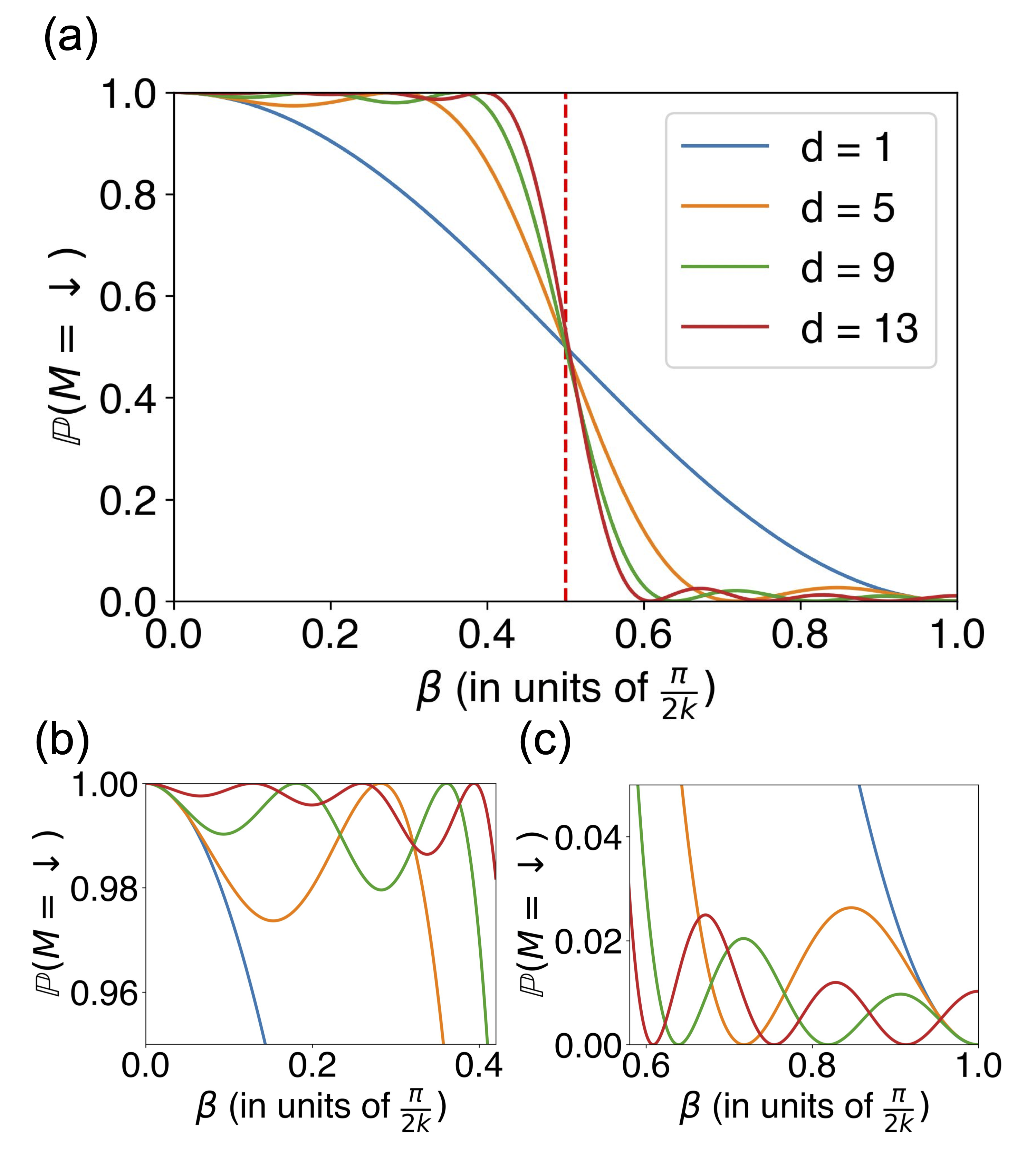}
        \caption{(a) Example response functions for various degrees $d$ for distinguishing a displacement with $\beta_{\rm th} = \frac{\pi}{4 \kappa}$, where $\kappa = 1 / 2048$, using QSPI phases from numerical optimization.
        (b, c) Magnified plots of the response function shown in (a) around $\mathbb{P}\left(M = \ \downarrow\right) = 0$ and $\mathbb{P}\left(M = \ \downarrow\right) = 1$, demonstrating that the response function for a $d$-QSPI protocol has $(d - 1)$ turning points in the interval $\left[0, \,\frac{\pi}{2\kappa}\right)$.
        }
        \label{fig:image14}
    \end{figure}

    Given the QSPI sensing protocol Protocol~\ref{pro:qspi-sensing}, we compute the corresponding response function based on converged optimization results for $\kappa = \frac{1}{2048}$, $\beta_{\rm th} = \frac{\pi}{4 \kappa}$ and plot this as a function of $\beta$ in Subfig.~\ref{fig:image14}(a) for degrees $d = $ 1, 5, 9, and 13, with Subfigs.~\ref{fig:image14}(b) and \ref{fig:image14}(c) zoomed in for the small and large $\beta$ regions, respectively. As shown in the figure, when $d = 1$, the response function simply takes the shape of a $\cos(\cdot)$ function. As the degree of the QSPI protocol increases, not only does the slope of the falling edge become steeper, but also, more ripples are observed in the wings, as highlighted in panels (b) and (c) of Fig.~\ref{fig:image14}. These ripples are a common feature for finite-degree polynomial approximations to discontinuous functions.

    Furthermore, a closer observation of the response function reveals that a $d$-QSPI protocol has precisely $(d - 1)$ local minima or maxima of its corresponding response function in the interval $\left[0, \,\frac{\pi}{2\kappa}\right)$. This is expected because a degree-$d$ polynomial on $\beta$ has at most $(d - 1)$ turning points. From a signal processing perspective, such qubit response functions serve as low-pass filters on the signal parameter $\beta$ \cite{oppenheim1999discrete}. We note that filters have been widely used in classical decision-making \cite{kuyu2016new}, where filter functions (or impulse response) of the desired shape can be implemented as infinitely smooth analytical functions (such as the Butterworth or Chebyshev filters) or as finite-degree polynomials. The former class of filters are often called infinite impulse response because an exact realization of the response would require an infinite order polynomial, while the latter are named finite impulse response. Our QSPI protocol can therefore be viewed as a quantum circuit realization of a finite impulse response (finite-degree filter) on a classical parameter $\beta$ which parameterizes a quantum process (the underlying displacement signal). There is efficient classical algorithm, the Parks-McClellan algorithm, that designs optimal finite-order polynomial filters \cite{parks1972chebyshev}. It is interesting to ask whether there might exist a quantum version of the Parks-McClellan algorithm for filter design. Another question is to what extent our QSPI protocol can realize classical filters of a given degree. Sufficient conditions provided in Theorem~\ref{thm:qsp-displacement-sensing} will shed more light on these questions.

\subsection{Scaling of Decision Quality}
\label{sec:numerical-heisenberg-scaling}

\begin{figure}[htbp]
\includegraphics[width=\columnwidth]{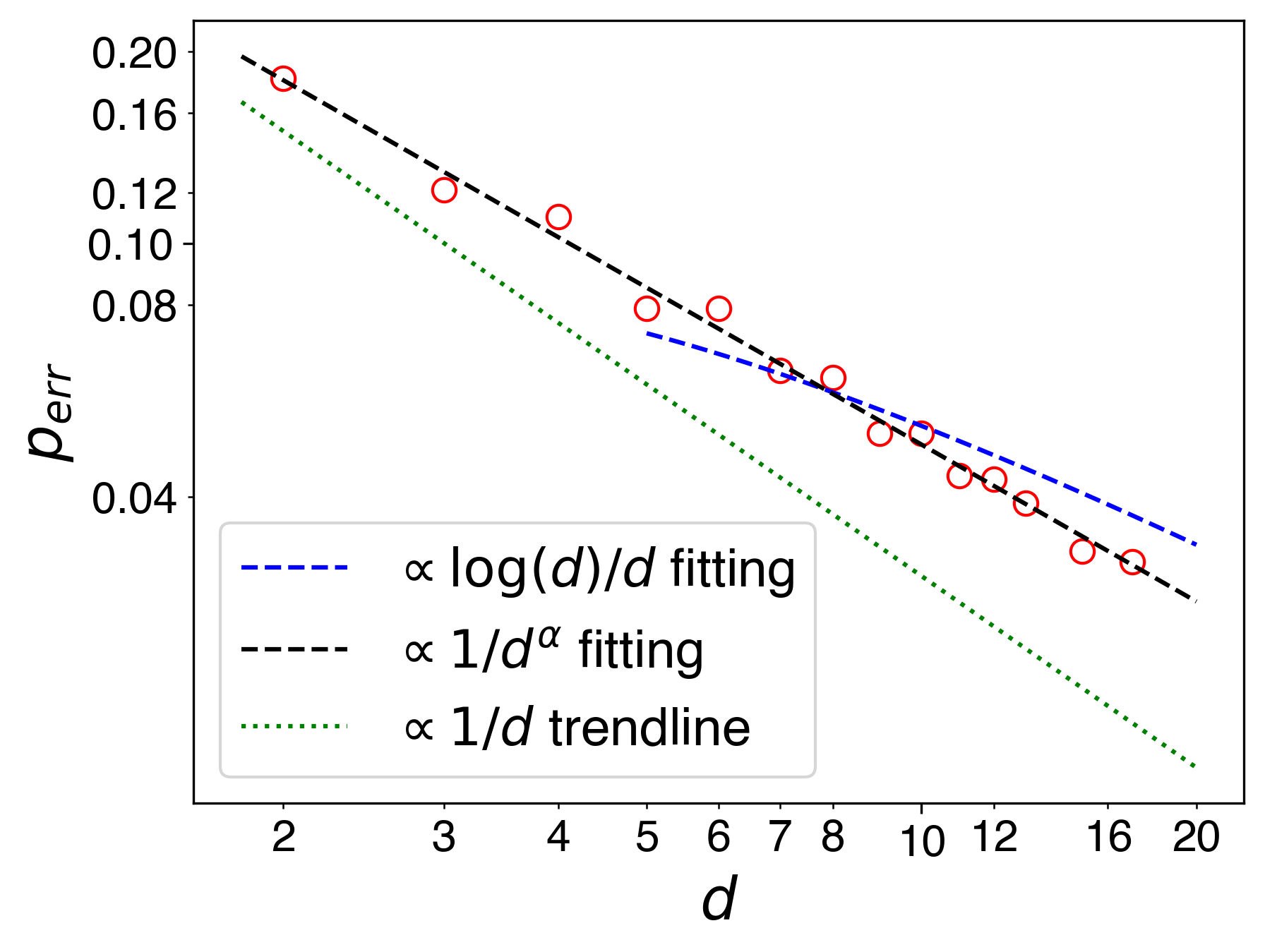}
\caption{Log-log scale plot of $p_{\rm err}$ versus the QSPI protocol degree $d$ (red circles). The best-fit power-law scaling (black dashed line) has a slope of $\alpha = -0.82 \pm 0.02$, indicating $p_{\rm err} \propto d^{-0.82 \pm 0.02}$. Best-fit theoretical scaling from Eq.~\eqref{perr_vs_d} using data for $d \ge 5$ is shown with a blue dashed line. The $\propto 1 / d$ efficient scaling is shown with a green dotted line for comparison.}
\label{fig:perr}
\end{figure}

To analyze our results and compare with the traditional displacement-sensing approaches, we plot the decision error of the QSPI protocol against its degree $d$ on a log-log plot, as shown in Fig.~\ref{fig:perr}. This plot illuminates an interesting relationship between the QSPI protocol degree and the associated response function.
From the figure, we can see that the numerical data points can be fitted by a linear black dashed line, demonstrating a power-law trend for $p_{\rm err}$ vs. degree $d$. The fitting reveals a slope of roughly $\alpha = -0.82 \pm 0.02$ which is close to the efficient scaling as defined in Definition \ref{def:hl-decision}. This power-law fit does not precisely fit all of the points, and it exhibits a parity-dependence with respect to the probability of error.
The slope is slightly worse than that for $1/d$ scaling (green dotted line), while still clearly outperforming the scaling expected for the standard quantum limit, a deficiency that is consistent with the additional logarithmic factor $\log(d)$, as explained in Eq.~\eqref{perr_vs_d} of Sec.~\ref{sec:binary-decision-making}.
To make a more direct comparison, we also fit all data points for $d \ge 5$ using the analytical scaling in Eq.~\eqref{perr_vs_d} (blue dashed line). The discrepancy between the analytical expectation and numerical data occurs because our analytical scaling is derived in the large $d$ limit, where we assume the major source of error $p_{\rm err}$ is from the falling edge, as seen in Fig.~\ref{fig:image14} (also see discussion in Appendix~\ref{app:heisenberg-scaling-proof}).

One final point is that although we were able to obtain numerical results and verify them, this brute-force optimization method does have several clear challenges. For example, to numerically optimize the sequence of phases minimizing the loss function ($p_{\rm err}$), one must select an initial value for each of the phases as a starting point.  It is possible that a sequence of initial values in the vicinity of a local minimum is selected and then the optimization procedure never escapes from the neighborhood around the local minimum. This problem becomes especially challenging for longer QSP sequences, where the search space is likely to contain more local minima. We attempt to address this difficulty by seeding with multiple distinct random initial phase sequences and iterating on only a subset with the least $p_{\rm err}$.  Thus, we are able to increase the likelihood that we find a phase sequence that has not been trapped in a local minimum, allowing us to discern the optimal scaling of the decision error with QSPI degree.

\subsection{Wigner Function of Optimal QSPI Sensing States}
\label{sec:wigner-function}

Given the numerically optimized phases for optimal QSPI sensing states, we may now visualize the resulting states to gain intuition about why their properties allow them to outperform cat states for our decision problems, as we describe in this section.  

\begin{figure*}[htb!]
    \centering
    \includegraphics[width = \textwidth]{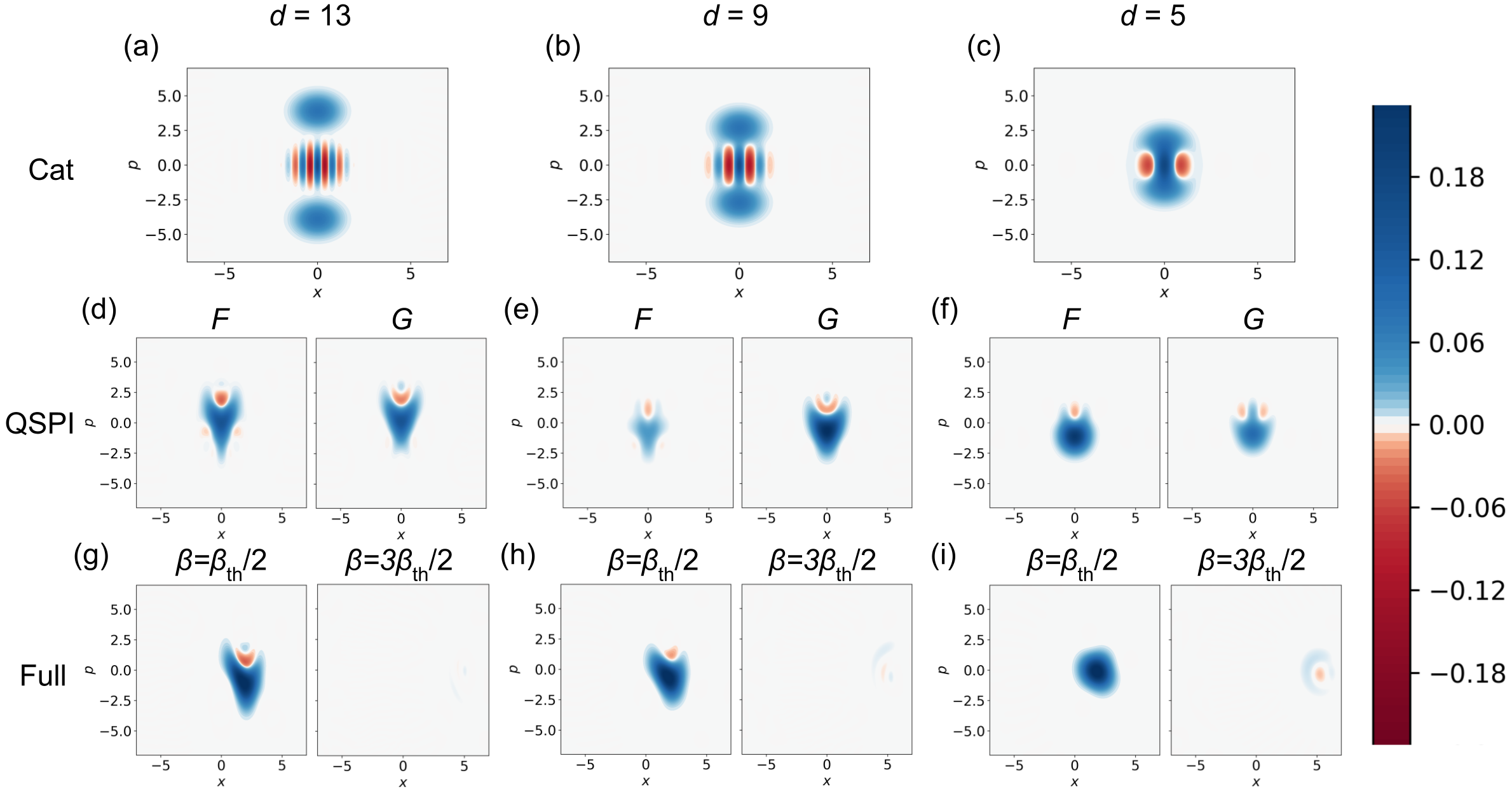}
    \caption{Wigner plots of $F$ for the cat state (Figs.~\ref{fig:wigner}(a), \ref{fig:wigner}(b), and \ref{fig:wigner}(c)), $F$ and $G$ for the optimal QSPI sensing state (Figs.~\ref{fig:wigner}(d), \ref{fig:wigner}(e), and \ref{fig:wigner}(f)), and the oscillator state resulting after the entire protocol is applied with displacements of $\beta = \frac{1}{2} \beta_{\rm th}$ and $\beta = \frac{3}{2} \beta_{\rm th}$ conditioned on the qubit being in the $\lvert \downarrow \rangle$ state (Figs.~\ref{fig:wigner}(g), \ref{fig:wigner}(h), and \ref{fig:wigner}(i)) with $\kappa = 0.15 \sqrt{2}$ and $\beta_{\rm th} = \frac{\pi}{4 \kappa} = \frac{5 \pi}{3 \sqrt{2}}$ constructed for $d = 13$, $9$, and $5$. Note the significant differences between the Wigner plots for the cat states and those for the QSPI sensing states, which do not closely resemble any known classes of quantum states.
    While the cat states all appear very similar but with more distance between their two coherent state parts and more interference fringes at the center as the degree $d$ increases, the optimal sensing states have a more complex interference pattern for improved decision-making.  As expected, the final oscillator state after the sensing protocol conditioned on the qubit being in the $\lvert \downarrow \rangle$ state represents a probability close to 1 for a displacement by $\beta = \frac{1}{2} \beta_{\rm th}$ but a small probability for a displacement by $\beta = \frac{3}{2} \beta_{\rm th}$, matching the behavior of the calculated response functions, shown for $\kappa = \frac{1}{2048}$ in Fig.~\ref{fig:image14}.  A symmetrical logarithmic scale, where the scaling is logarithmic in both the positive and negative directions from a small linearly-scaling range around zero, is used as the color scheme in order to increase the contrast of finer features of the Wigner quasiprobability distribution.
    }
    \label{fig:wigner}
\end{figure*}

Although we learned the QSPI phases for the small $\kappa = 1 / 2048$ in order to best decouple the $c_s$ coefficients and hence facilitate the optimization of the phases, bear in mind the discussion of the coupling of $\kappa$ and $\beta$ in Eq.~\eqref{cs-def-app} and realize that we can adjust our choice of $\kappa$ and the corresponding $\beta$ with only minimal fine-tuning of the phases learned for the original value of $\kappa$. As such, for clearer visualization of low-degree states on the Wigner plots, we increase the scale of our problem by setting $\kappa = 0.15 \sqrt{2}$ and setting $\beta_{\rm th} = \frac{\pi}{4 \kappa} = \frac{5 \pi}{3 \sqrt{2}}$.  For each degree $d$, we use the phase sequence learned for $\kappa = \frac{1}{2048}$ as our initial phase sequence and resume optimization until convergence for the new value of $\kappa$. This change to $\kappa$ results in minimal change to the QSPI phases during the optimization, with the majority of them differing by less than 1\% relative to their original values. The Wigner plots for $F$ and $G$ (Eq.~\eqref{eq:fwgw-poly}) for the newly optimized states resulting from these $d$-QSPI protocols with $\kappa = 0.15 \sqrt{2}$ are shown for $d = 5$, $9$, and $13$ in Subfigs.~\ref{fig:wigner}(d) -- \ref{fig:wigner}(i) in the lower half of Fig.~\ref{fig:wigner}.

To compare these QSPI states with cat states, we plot $F$ (Eq.~\eqref{eq:fw-poly}) for cat states constructed with the same number $d - 1$ of displacements by $\kappa$ in Subfigs.~\ref{fig:wigner}(a) -- \ref{fig:wigner}(c) in the upper panel of Fig.~\ref{fig:wigner}.  We see immediately that the interference patterns (regions with large contrast and Wigner negativity) for the cat state and QSPI state of same degree $d$ differ significantly. The cat states display interference fringes from their two displaced coherent states that oscillate with a frequency of $\sqrt{2}d\kappa/\pi$ along the $x$-axis. Thus, cat states corresponding to higher degree $d$ have higher-frequency fringes; this is what makes cat states effective sensors of very small displacements perpendicular to the coherent-state displacement in parameter-estimation protocols \cite{gilmore2021quantum} but not optimal for making global decisions. In fact, cat states have a constant probability of error $p_{\rm err}$ across all $d$ for a given $\beta_{\rm th}$ and $\kappa$, as explained in Sec.~\ref{sec:cat-state-limitation-dec}. 

This trade-off between high and low frequency features appears in Bayesian approaches to estimation problems as well since without prior information, one must consider a flat prior over some fixed region and resolve ambiguities at the cost of devoting resources to low frequency features \cite{G_recki_2020,Berry_2000}. Often, probe states achieving Heisenberg scaling are optimal for local estimation, but only in a small neighborhood around the true value as we see for the cat state interferometry. If there is no prior information about which small neighborhood of the response function is being examined, then there is \emph{fringe ambiguity}. Thus, the decision problem we address is akin to resolving fringe ambiguity issues in parameter estimation, and our QSPI response function provides a way of interpolating between local and global estimation regimes.  For example, in interferometric phase estimation one must restrict their prior to an interval $2\pi/(\lambda_+ - \lambda_-)$, where $\lambda_{\pm}$'s are extreme eigenvalues of generator related to signal oracle, as phases differing by a multiple of $2\pi/(\lambda_+ - \lambda_-)$ cannot be distinguished without additional resources \cite{G_recki_2020}. This trade off between local and global estimation is also pointed out in \cite{meyer2023quantum, marciniak_optimal_2022} where the short period of probe state dynamics in $n$-qubit GHZ states offer optimal local estimation in a range $2\pi/n$ but cannot distinguish parameter values differing by a multiple of $2\pi/n$. Since the spacing between ambiguous fringes is known for such problems, one could design efficient low-degree quantum filters that discriminate only as precisely as necessary by choosing the optimal response function for the given spacing.

The physical intuition for this improved decision-making of the QSPI states stems from the lower-frequency features shown in the Wigner contrast plot. The spacing between sharp-Wigner-contrast features which appear consistently in the upper half of plots in Subfigs.~\ref{fig:wigner}(d) and \ref{fig:wigner}(f) around $(x \approx \pm 1, \,p \approx 1)$ indicates the placement of sharp thresholds
in the response function. This is why the cat-state response function places sharp thresholds with frequency $d\kappa/\pi$. In comparison, for general QSPI states, the interplay between asymmetric $F$ and $G$ is more complicated to analyze than for the $F$ and $G$ of cat states, but as we see in Fig.~\ref{fig:wigner}, the spacings between sharp Wigner features shown in Subfigs.~\ref{fig:wigner}(d) and \ref{fig:wigner}(e) remain larger than those of the cat states shown in Subfigs.~\ref{fig:wigner}(a) and \ref{fig:wigner}(b), respectively. 
Notice that the Wigner extent is also smaller for QSPI states as compared with that of cat states. This is because cat states are maximally extended in phase space for the given energy of the protocol by na\"{i}vely shifting the wave packet along one direction, while the QSPI state devotes some of the energy to creating a more complicated and optimized interference pattern phase space as compared to the simple sinusoid with Gaussian envelope that the cat state creates.

Subfigs.~\ref{fig:wigner}(g) -- \ref{fig:wigner}(i) then depict the final oscillator state after performing the entire QSPI displacement-sensing protocol with displacements by $\beta = \frac{1}{2} \beta_{\rm th}$ and $\beta = \frac{3}{2} \beta_{\rm th}$, conditioned on the qubit being in the $\lvert \downarrow \rangle$ state.  We note from these subfigures that the quasiprobability distribution exhibits primarily constructive interference to give a total $\lvert \downarrow \rangle$-measurement probability of nearly 1 when the QSPI sensing protocol is applied for a displacement by $\beta = \frac{1}{2}\beta_{\rm th}$, while the Wigner quasiprobability distribution exhibits primarily destructive interference to give a total $\lvert \downarrow \rangle$-measurement probability of nearly 0 when the QSPI sensing protocol is applied for a displacement by $\beta = \frac{3}{2} \beta_{\rm th}$ (hence the nearly empty Wigner plots). These behaviors agree with our theoretical analysis and solve the Main Problem stated in Sec.~\ref{sec:quantum-decision-def}.

\begin{table*}
    \centering
    \begin{tabular}{c|c|c} \hline\hline
        Sensing State & $\mathbb{P}(M = \ \downarrow | \beta = \frac{1}{2} \beta_{\rm th})$ & $\mathbb{P}(M = \ \downarrow | \beta = \frac{3}{2} \beta_{\rm th})$ \\ \hline
        Cat State & 0.957 & 0.035 \\ \hline
        QSPI 5 & 0.956 & 0.035 \\ \hline
        QSPI 9 & 0.976 & 0.021 \\ \hline
        QSPI 13 & 0.982 & 0.016 \\ \hline\hline
    \end{tabular}
    \caption{\label{tab:probs} The probability of measuring the qubit in the $\lvert \downarrow \rangle$ state after applying the entire sensing protocol with displacements by $\frac{1}{2} \beta_{\rm th}$ (below threshold) and $\frac{3}{2} \beta_{\rm th}$ (above threshold) using the cat state (independent of degree $d$) and the $d$-QSPI states for $d = $ 5, 9, and 13, where $\kappa = 0.15 \sqrt{2}$ and $\beta_{\rm th} = \frac{\pi}{4 \kappa} = \frac{5 \pi}{3 \sqrt{2}}$. These values are calculated from a numerical simulation of the QSPI protocol with a Fock-level truncation of $N = 500$ and using a grid with a unit cell size of $0.2 \times 0.2$. The numbers in the table are confirmed to converge to the given significant figures with respect to both Fock-level truncation $N$ and grid spacing by performing the same calculations with larger $N$ and finer grids.
    }
\end{table*}

In particular, we also show in Table~\ref{tab:probs} the probabilities of measuring $\lvert \downarrow \rangle$ for the qubit state after performing the entire protocol with both the $d$-QSPI sensing states for $d = $ 5, 9, and 13 and the corresponding cat states for reference.  Note that according to simulation results (and as predicted in Sec.~\ref{sec:cat-state-limitation-dec}), the probability of detection does not change with degree $d$ for the cat state, so we provide only one probability.  Note also from the results shown in the table that although the cat state performs well for sensing a displacement by $\beta_{\rm th}$ (with $\kappa = 0.15 \sqrt{2}$ and $\beta_{\rm th} = \frac{\pi}{4 \kappa} = \frac{5 \pi}{3 \sqrt{2}}$), its performance is already matched with only a 5-QSPI state.  Moreover, while the cat state's displacement-sensing performance remains constant with increasing degree $d$, the $d$-QSPI state's performance improves, so the 9- and 13-QSPI states outperform the cat state; moreover, performance of $d$-QSPI states for this displacement-sensing task will continue to improve as $d$ increases further.

\section{Conclusion}
\label{sec:conclusion}

In this paper, we present a general framework for single-shot quantum sensing using continuous-variable systems by establishing a theory of quantum signal processing interferometry. The basics of this construction are a generalization of QSP to systems with a qubit coupled to a quantum harmonic oscillator. The controlled displacement operation between the qubit and the oscillator forms a natural block-encoding of the displacement operator on the oscillators, via which arbitrary polynomial transformations on the oscillator's quadrature operator can be efficiently implemented. The flexibility of QSP provides the basis for our algorithmic QSPI sensing protocol. A measurement on the qubit induces a qubit response function that is a polynomial transformation of the signal parameters that we would like to sense. By tuning the QSPI phase angles to design appropriate response functions, useful information about the signal parameters can be extracted efficiently.

The QSPI sensing protocol is analyzed in detail for a binary decision problem on a displacement channel with theoretical bounds on the sensing-circuit and sampling complexity. These binary decision oracles are then used to construct a composite protocol for parameter estimation via classical binary search and majority vote.  
Our sensing scheme is applied to determine if a displacement on the oscillator is greater or smaller than a certain threshold, and efficient scaling behavior is derived analytically and observed numerically for this application.

While we have demonstrated efficient scaling for a binary decision on a displacement channel, the sensing protocol can be further improved. One immediate task is to determine if there exist non-optimization-based algorithms for finding QSPI phase angles that achieve a general QSPI response function. This goal implies the need to find not only necessary but also sufficient conditions on QSPI for the backward direction of Theorem~\ref{thm:qsp-displacement-sensing}. Moreover, the sensing protocol for a single canonical variable of the oscillator can be generalized to two conjugate canonical variables of position and momentum simultaneously, allowing for the realization of quantum sensing in the entirety of phase space for the oscillator. Due to the Heisenberg uncertainty principle, tradeoffs between sensitivity in the position and momentum quadratures may be imposed using squeezing operations depending on the particular sensing task. In addition, the sensing power can be further enhanced by coherently manipulating multiple bosonic modes \cite{zhuang_distributed_2018,kwon2022quantum}, which can likely be coupled together with beamsplitters. In this context, tradeoffs between available quantum resources, such as space (number of oscillators) and time (sensing circuit depth), would be interesting to investigate.

The algorithmic QSPI-based quantum sensing protocol presented in this work opens many possibilities for useful applications. For example, bosonic modes appear in many physical systems, such as in molecular vibrations and light-matter interactions under confined conditions. The displacement-sensing scheme presented here can be used to sense any chemical environment change in molecules, as long as the change leads to an effective displacement operation on molecular vibrations, as in the case of ro-vibrational coupling \cite{pekeris1934rotation}, or a displacement on photonic modes, as in polariton chemistry \cite{ribeiro2018polariton,xiong2023molecular}. The flexibility of designing response functions can be used to deal with situations where the underlying signal has some prior distribution. Our framework also can be used to study the few-shot regime and connect local and global estimation strategies as in \cite{meyer2023quantum}; we show a new perspective for solving such metrology problems by focusing on response function filter design (as opposed to optimization over more abstract POVMs).

We note that the decomposition of composite sensing tasks, such as parameter estimation, into a series of decision problems provides ample room for the incorporation of hybrid quantum-classical algorithms into the sensing framework. For example, sophisticated adaptive strategies can be built in to gradually change the precision and shape of the decision filter and hence reduce the sensing cost. Lastly, sensing is prone to quantum noise. It would be useful to analyze the stability of our QSPI-based continuous-variable sensing protocol in the presence of quantum noise \cite{dong_beyond_2022,xu2023quantum}. Realistic noise models on hardware containing bosonic degrees of freedom could be incorporated into numerical simulations as well \cite{brady2023advances}. Both photon loss and heating (common to superconducting and trapped ion hardware, respectively) will change behavior of the filter if quantum error-correction is not incorporated into the QSP sequences \cite{demkowicz2015quantum,escher2011general,demkowicz2012elusive}. A future direction of interest would be to study the behavior of QSPI under noisy conditions or to extend the QSPI protocol into an error-corrected version \cite{zhou2019error}.

\begin{acknowledgments}
    GM and JSS were supported in part by the Army Research Office under the CVQC project W911NF-17-1-0481. GM, JSS, and YL were supported in part by NTT Research. The work of YL and ILC on analysis and numerical simulation was supported by the U.S. Department of Energy, Office of Science, National Quantum Information Science Research Centers, Co-design Center for Quantum Advantage (C$^2$QA) under contract number DE-SC0012704. The numerical simulations were performed with computing resources from subMIT at MIT Physics and the MIT SuperCloud/Lincoln Laboratory Supercomputing Center \cite{reuther2018interactive}, to whom we are grateful.
\end{acknowledgments}

\appendix 


\section{Proof of Bosonic QSP Theorem~\ref{thm:bosonic-qsp}}
\label{app:bosonic-qsp-proof}

Denote the set of eigenvalues and eigenvectors of the generator $h(\hat{x}, \,\hat{p})$ in Lemma~\ref{def:qubitization-mode} to be $\lambda$ and $\ket{\lambda}$ such that
\begin{align}
    h(\hat{x}, \,\hat{p}) \ket{\lambda} = \lambda \ket{\lambda}.
\end{align}
Since $h(\hat{x}, \,\hat{p})$ is Hermitian, we have $\lambda \in \mathbb{R}$. Moreover, for an infinite-dimensional oscillator, $\{ \lambda \}$ can be inherently continuous. Furthermore, we assume that $\{ \ket{\lambda} \}$ forms a (over)complete basis for the oscillator (for example, in the case of a continuous displacement operator whose eigenstates are coherent states) for achieving universal control of the oscillator. However, the bosonic QSP formalism still works in the subspace expanded by $\{ \ket{\lambda} \}$.

Now, consider the action of $W_z$ on an arbitrary qubit-oscillator entangled state, where the oscillator state is given by $\frac{1}{\sqrt{2}}(\ket{0} \ket{\Psi_0}_{\rm osc} + \ket{1} \ket{\Psi_1}_{\rm osc})$. Expand the oscillator state under the $\{ \ket{\lambda} \}$ basis to find
\begin{align}
    & W_z \frac{1}{\sqrt{2}} \left(\ket{0} \ket{\Psi_0}_{\rm osc} + \ket{1} \ket{\Psi_1}_{\rm osc} \right) \nn \\
    =& e^{-ih(\hat{x}, \,\hat{p}) \hat{\sigma}_z} \frac{1}{\sqrt{2}} \left(\ket{0} \ket{\Psi_0}_{\rm osc} + \ket{1} \ket{\Psi_1}_{\rm osc} \right) \nn \\
    =& e^{-i h(\hat{x}, \,\hat{p}) \hat{\sigma}_z} \frac{1}{\sqrt{2}} \left(\ket{0} \int d\lambda c_{0, \,\lambda} \ket{\lambda} + \ket{1} \int d\lambda c_{1, \,\lambda} \ket{\lambda} \right) \nn \\
    =& \frac{1}{\sqrt{2}} \left(\ket{0} \int d\lambda e^{-i\lambda}  c_{0, \,\lambda} \ket{\lambda} + \ket{1} \int d \lambda e^{i \lambda}  c_{1, \,\lambda} \ket{\lambda} \right) \nn \\
    =& \int d \lambda \ket{\lambda}  \frac{1}{\sqrt{2}} \left(\ket{0} e^{-i \lambda}  c_{0, \,\lambda} + \ket{1} e^{i\lambda}  c_{1, \,\lambda} \right) \nn \\
    =& \int d\lambda \ket{\lambda} \otimes e^{-i \lambda \sigma_z} \left[ \frac{c_{0, \,\lambda} \ket{0} + c_{1, \,\lambda} \ket{1} }{\sqrt{2}} \right].
    \label{wz_effect_subspace}
\end{align}
Therefore, $W_z$ acts individually on each $2 \times 2$ subspace labeled by the eigenvalue $\lambda$ of $h(\hat{x}, \,\hat{p})$ in a similar spirit to that of qubitization of a finite-dimensional block-encoding. From Eq.~\eqref{wz_effect_subspace}, repeatedly applying $W_z$ and a single-qubit rotation $e^{i \theta_j \hat{\sigma}_x}$ will result in the application of QSP to each individual $2 \times 2$ subspace:
\begin{align}
    & e^{i \theta_d \hat{\sigma}_x} \prod_{j = 0}^{d - 1} W_z e^{i \theta_j \hat{\sigma}_x} \frac{1}{\sqrt{2}}(\ket{0} \ket{\Psi_0}_{\rm osc} + \ket{1} \ket{\Psi_1}_{\rm osc}) \nn \\
    =& \int d\lambda \ket{\lambda} \otimes e^{i\theta_d \hat{\sigma}_x} \prod_{j = 0}^{d - 1}  e^{-i \lambda \hat{\sigma}_z} e^{i \theta_j \hat{\sigma}_x}  \left[ \frac{c_{0, \,\lambda} \ket{0} + c_{1, \,\lambda} \ket{1} }{\sqrt{2}} \right].
\end{align}
In the $2 \times 2$ subspace for each $\lambda$, we identify the following sequence of $SU(2)$ rotations
\begin{align}
    U_{\lambda, \,\vec{\theta}} = e^{i\theta_d \hat{\sigma}_x} \prod_{j = 0}^{d - 1}  e^{-i \lambda \hat{\sigma}_z} e^{i \theta_j \hat{\sigma}_x}.
\end{align}
This is nothing but the usual single-qubit QSP sequence under the $W_z$-convention \cite{MRTC21}, where the signal being transform is a Pauli-$Z$ rotation parameterized by the eigenvalue $\lambda$. To be more concrete, applying the single-qubit QSP theorem to $U_{\lambda, \,\vec{\theta}}$, we have
\begin{align}
    U_{\lambda, \,\vec{\theta}} = 
    \begin{bmatrix}
        F(\omega_\lambda) & i G(\omega_\lambda) \\
        i G(\omega_\lambda^{-1}) & F(\omega_\lambda^{-1})
    \end{bmatrix},
\end{align}
where $\omega_\lambda = e^{-i \lambda}$, $F(\omega_\lambda) = \sum_{j = -d}^d f_j \omega_\lambda^j$, $G(\omega_\lambda) = \sum_{j = -d}^d g_j \omega_\lambda^j$ are Laurent polynomials of degree $d$ with real coefficients $f_j, \,g_j \in \mathbb{R}$. The unitarity condition on $U_{\lambda, \,\vec{\theta}}$ also requires that 
\begin{align}
    F(\omega_\lambda) F(\omega_\lambda^{-1}) + G(\omega_\lambda) G(\omega_\lambda^{-1}) = 1,
    \label{eq:f-g-appendix}
\end{align}
$\forall ~\lambda$. The reverse direction of the QSP theorem also means that given arbitrary degree-$d$ real Laurent polynomial $F(\cdot), \,G(\cdot)$ satisfying Eq.~\eqref{eq:f-g-appendix}, there exists a set of phase angle $\vec{\theta}$ such that a circuit constructed from $U_{\lambda, \,\vec{\theta}}$ can implement the given $F(\cdot), \,G(\cdot)$. 

The above single-qubit QSP applies for each individual eigenspace of the oscillator labeled by $\lambda$. Performing the integral over all $\lambda$, it follows that the overall sequence in Eq.~\eqref{qsp_theta} performs a Laurent polynomial transformation on $e^{-i h(\hat{x}, \,\hat{p})}$, hence proving Theorem~\ref{thm:bosonic-qsp}. The last step of elevating from qubit QSP to the hybrid qubit-oscillator continuous-variable case resembles the spirit of deriving quantum eigenvalue transform \cite{MRTC21}, with the difference that in our case the spectrum of the oscillator is continuous, while in the usual multi-qubit case, the eigen spectrum is discrete. 

\section{Proof that Response Function is Polynomial of Sensed Signal}
\label{app:perr-real-proof}

We prove in this section that the QSPI response function, as defined in Eq.~\eqref{p00_def}, is a degree-$d$ polynomial transformation of the new signal $\nu = e^{i (2 \kappa) \beta }$ within a restricted range $\left[-\frac{\pi}{2 \kappa}, \,\frac{\pi}{2 \kappa}\right]$ and the polynomial is real. 

Using the Laurent polynomial expressions in Eq.~\eqref{eq:fwgw-poly} and explicitly evaluating the integration with respect to $x$, we can write Eq.~\eqref{p00_def} as a series sum
\begin{align}
    \mathbb{P}(M = \ \downarrow | \beta)  = \sum_{n, \,n', \,m, \,m' = -d}^d  A_{n, \,n', \,m, \,m'},
    \label{prob-def-app}
\end{align}
where 
\begin{align}
    A_{n, \,n', \,m, \,m'} &= (f_n f_{n'} + g_n g_{n'})(f_m f_{m'} + g_m g_{m'})^*  \nonumber \\
     & \times e^{-\frac{1}{4} \kappa^2 (n - n' - m + m')^2} e^{-i \kappa (n - m) \beta} .
     \label{Annmm-def}
\end{align}
The following property can be verified:
\begin{align}
    A_{n, \,n', \,m, \,m'} = A^*_{m, \,m', \,n, \,n'}.
\end{align}
Therefore, we can rearrange the sum for the QSPI response function to be
\begin{widetext}
    \begin{align}
        \mathbb{P}(M = \ \downarrow|\beta) 
        &= \frac{1}{2} \sum_{n, \,n', \,m, \,m' = -d}^d  ( A_{n, \,n', \,m, \,m'} + A^*_{m, \,m', \,n, \,n'})\nonumber \\
        &= \frac{1}{2} \left( \sum_{n, \,n', \,m, \,m' = -d}^d A_{n, \,n', \,m, \,m'} + \sum_{n, \,n', \,m, \,m' = -d}^d A^*_{m, \,m', \,n, \,n'} \right) \nonumber \\
        &= \frac{1}{2} \left( \sum_{n, \,n', \,m, \,m' = -d}^d A_{n, \,n', \,m, \,m'} + \sum_{n, \,n', \,m, \,m' = -d}^d A^*_{n, \,n', \,m, \,m'} \right) \nonumber \\
        &= \sum_{n, \,n', \,m, \,m' = -d}^d \Re{A_{n, \,n', \,m, \,m'}},
    \end{align}
\end{widetext}
where we have renamed variables $m \leftrightarrow n$ and $m' \leftrightarrow n'$ from the second line to the third line, and $\Re{\cdot}$ denotes the real part.

Next, we prove that $\mathbb{P}(M = \ \downarrow | \beta)$ is a degree-$d$ Laurent polynomial of $\nu = e^{i (2 \kappa) \beta}$ in the range $\left[-\frac{\pi}{2 \kappa}, \,\frac{\pi}{2 \kappa}\right]$. 

Because $f_n, \,g_n$ are each coefficients of Laurent polynomials from QSP, it follows that $f_n, \,g_n \ne 0$ only for even $n$ if $d$ is even, or $f_n, \,g_n \ne 0$ only for odd $n$ if $d$ is odd. This means $A_{n, \,n', \,m, \,m'} \ne 0$ only when $m, \,n$ have the same parity and $m', \,n'$ have the same parity, which further suggests the variable substitution
\begin{align}
    m = n + 2s, ~~~ m' = n' + 2r,
\end{align}
where $-d \le s, \,r \le d$. 
Substituting this back into Eq.~\eqref{Annmm-def}, we have
\begin{align}
    A_{n, \,n', \,n + 2s, \,n' + 2r} &= (f_n f_{n'} + g_n g_{n'}) \nonumber \\
    & \quad \quad \times (f_{n + 2s} f_{n' + 2r} + g_{n + 2s} g_{n' + 2r})  \nonumber \\
     & \quad \quad  \times e^{- \kappa^2 (r - s)^2} e^{i (2 \kappa) s \beta}.
     \label{Annmm-def-new}
\end{align}
Further, substituting this back into Eq.~\eqref{prob-def-app}, we obtain
\begin{align}
       \mathbb{P}(M = \ \downarrow | \beta)  
     =  \sum_{s = -d}^{d} c_s(\kappa) \nu^s,
\end{align}
where $\nu = e^{i (2 \kappa) \beta}$ and
\begin{align}
      c_s(\kappa) &= \sum_{n, \,n', \,r= -d}^d (f_n f_{n'} + g_n g_{n'}) \nonumber \\
    & \quad \quad \quad \times (f_{n + 2s} f_{n' + 2r} + g_{n + 2s} g_{n' + 2r}) e^{- \kappa^2 (r - s)^2},
    \label{cs-def-app}
\end{align}
with $f_n, \,g_n \in \mathbb{R}$, $c_s \in \mathbb{R}$.

Because the new signal operator $\nu$ has an effective momentum of $2 \kappa$, this means that $\mathbb{P}(M = \ \downarrow | \beta)$ will be periodic with a reduced period of $\left[-\frac{\pi}{2 \kappa}, \,\frac{\pi}{2 \kappa}\right]$. It follows that the QSPI response function $\mathbb{P}(M = \ \downarrow|\beta)$ is a degree-$d$ Laurent polynomial in the operator $\nu = e^{i (2 \kappa) \beta}$.

\section{Recursive Relationship Between QSP Coefficients}
\label{app:qsp-coeff-recursive}

The probability of making the wrong decision can be efficiently computed classically from the original QSP phase angles. First, by using the following recursive relationship, all the QSP coefficients $f_n, \,g_n$ as stated in Theorem \ref{thm:bosonic-qsp} can be computed from the phase angles. Second, the series sum in Eq.~\eqref{perr-analytical} can be evaluated explicitly using the computed $f_n, \,g_n$, without loss of numerical precision.
\begin{align}
    f_r^{(d + 1)} &= 
    \begin{cases}
        \cos\theta_{d + 1} f_{r - 1}^{(d)}, ~~ r = d, \,d + 1 \\
        -\sin\theta_{d + 1} g_{r + 1}^{(d)}, ~~ r = -d, \,-d - 1 \\
        \cos\theta_{d + 1} f_{r - 1}^{(d)} -\sin\theta_{d + 1} g_{r + 1}^{(d)}, ~~ |r| \le (d - 1)
    \end{cases}\\
    g_r^{(d + 1)} &= 
    \begin{cases}
        \sin\theta_{d + 1} f_{r - 1}^{(d)}, ~~ r = d, \,d+1 \\
        \cos\theta_{d + 1} g_{r + 1}^{(d)}, ~~ r = -d, \,-d-1 \\
        \sin\theta_{d + 1} f_{r - 1}^{(d)} + \cos\theta_{d + 1} g_{r + 1}^{(d)},  ~~ |r| \le (d - 1)
    \end{cases}
\end{align}

\section{Proof of Efficient Scaling for QSPI Binary Decisions}
\label{app:heisenberg-scaling-proof}

From Fig.~\ref{fig:perr-schematic}, the decision error probability can be approximately written as a sum of two contributions in the following way
\begin{align}
    p_{\rm err} \approx \epsilon \left(\frac{\pi}{\kappa} - 2\sigma \right) + \frac{\sigma}{2},
    \label{perr_from_polyerr}
\end{align}
where $\epsilon$ is the approximation error to an ideal step function from a polynomial function. The first term in Eq.~\eqref{perr_from_polyerr} is obtained from such an imperfect polynomial approximation in the region of $\left[-\frac{\pi}{2 \kappa}, \,\frac{\pi}{2 \kappa}\right]$, excluding the rising and falling edges $\left[\beta_{\rm th} - \sigma / 2, \,\beta_{\rm th} + \sigma / 2\right] \cup \left[-\beta_{\rm th} - \sigma / 2, \,-\beta_{\rm th} + \sigma / 2\right]$; the second term in Eq.~\eqref{perr_from_polyerr} is from erroneous decisions when the displacement $\beta$ lies within the rising or falling edge. 

Rearranging Eq.~\eqref{perr_from_polyerr}, this means that the error in the polynomial approximation to $P_{\rm ideal}^{\rm sign, \,sin}$ is
\begin{align}
    \epsilon \approx \frac{p_{\rm err} - \frac{\sigma}{2}}{\frac{\pi}{\kappa} - 2 \sigma}.
    \label{polyerr-from-perr}
\end{align}
From Ref.~\cite{martyn2023efficient}, to achieve an $\epsilon$ approximation to the sign function in regions excluding $\left[-\sigma/2, \,\sigma/2\right]$ requires a polynomial of degree $d = \gamma(\epsilon, \,\sigma)$ for
\begin{widetext}
    \begin{align}
        \gamma(\epsilon, \,\sigma) := 2 \cdot \Bigg\lceil \text{max}\left(\frac{e}{\sigma} \sqrt{W\left(\frac{8}{\pi \epsilon^2} \right) W\left(\frac{512}{e^2 \pi} \frac{1}{\epsilon^2} \right)}, \,  \sqrt{2} W\left( \frac{8\sqrt{2}}{\sqrt{\pi} \sigma \epsilon} \sqrt{W\left(\frac{8}{\pi \epsilon^2} \right)} \right) \right) \Bigg\rceil + 1,
        \label{gamma-def}
\end{align}
\end{widetext}
where $W(\cdot)$ is the Lambert W function. Assuming $\sigma = O(p_{\rm err})$ and for small $p_{\rm err}$, substitute Eq.~\eqref{polyerr-from-perr} into Eq.~\eqref{gamma-def} and use a Taylor expansion on the Lambert W function in order to obtain Eq.~\eqref{d_vs_perr} in the main text.

\bibliographystyle{quantum}
\bibliography{ref}

\end{document}